\documentclass[preprint,floatfix, showpacs,superscriptaddress]{revtex4}
\usepackage{color,graphicx}
\usepackage{bm}
\usepackage{amsmath}
\usepackage{amssymb}
\usepackage{float}

\usepackage{changes}

\newcommand{\be}{\begin{eqnarray}}
\newcommand{\ee}{\end{eqnarray}}

\begin{document}
\title{Local Quantum Criticality in the Two-dimensional  Dissipative Quantum XY Model}
\author{Lijun Zhu}
\affiliation{Department of Physics and Astronomy, University of California, Riverside CA 92521, USA}
\author{Yan Chen}
\affiliation{Department of Physics and State Key Laboratory of Surface Physics, Fudan University, Shanghai 200433, China}
\author{Chandra M. Varma}
\affiliation{Department of Physics and Astronomy, University of California, Riverside CA 92521, USA}

\begin{abstract}
We use quantum Monte-Carlo simulations to  calculate the phase diagram and the correlation functions for the quantum phase transitions in the two-dimensional dissipative quantum XY model with and without four-fold anisotropy. Without anisotropy, the model describes the superconductor to insulator transition in two-dimensional dirty superconductors. With anisotropy, the model represents the loop-current order observed in the under-doped cuprates and its fluctuations, as well as the fluctuations near the ordering vector in simple models of two-dimensional itinerant ferromagnets and itinerant antiferromagnets. These calculations test an analytic solution of the model which re-expressed it in terms of topological excitations - the vortices with interactions only in space but none in time, and warps with leading interactions only in time but none in space, as well as sub-leading interactions which are both space and time-dependent. For parameters where the proliferation of warps dominates the phase transition, the critical fluctuations as functions of the deviation of the dissipation parameter $\alpha$ on the disordered side from its critical value $\alpha_c$ are scale-invariant in imaginary time $\tau$ as the correlation length in time $\xi_{\tau} = \tau_c e^{[\alpha_c/(\alpha_c-\alpha)]^{1/2}}$ diverges, where $\tau_c$ is a short time cut-off. On the other hand, the spatial correlations develop  with a correlation length $\xi_x \approx \xi_0 \log{(\xi_{\tau})}$, with $\xi_0$ of the order of a lattice constant. The dynamic correlation exponent $z$ is therefore $\infty$. The Monte-Carlo calculations also directly show properties of warps and vortices. Their densities and correlations across the various transitions in the model are calculated and related to those of the order parameter correlations in the dissipative quantum XY model.
\end{abstract}
\maketitle

\section{Introduction}
The dissipative quantum XY model was introduced \cite{chakra86, Fisher86} to describe the observed quantum phase transition \cite{expts} in thin metallic films from a superconductor to insulator at a universal value of their normal state resistance. In the past few years, the model has acquired applications in other physical contexts.  The new physical contexts are the quantum-critical point of the loop-current order in cuprate superconductors \cite{cmv, bourges, Aji2007}, and the 2D itinerant antiferromagnetic(AFM) quantum-critical point \cite{cmv-afm}, which may be of relevance to Fe-based superconductors and some heavy fermion compounds. It is also of course directly applicable to quantum critical points in 2D XY ferromagnets.

Electronic-fluctuation induced superconductivity in cuprates, Fe-based superconductors, and in heavy-fermion superconductors always appears together with a normal state which does not obey the Fermi-liquid paradigm. For cuprates, the properties of the normal state, sometimes called the strange metal phase, could be phenomenologically described as a marginal Fermi liquid (MFL) \cite{Varma1989}, in which the coupling of electrons is to quantum-critical fluctuations which are local in space and power-law in time.  Such critical fluctuations violate the paradigm of classical dynamical critical fluctuations \cite{Hohenberg} or their simple quantum analogs \cite{Beal-Monod, Hertz, Moriya}. The anomalous normal-state properties have been associated with quantum criticality of an order competing with superconductivity.  Thermodynamic and transport properties near the antiferromagnetic quantum critical point of some heavy fermion compounds \cite{Lohneysen-rev} and at least some of the Fe-based superconductors near their antiferromagnetic quantum critical point are remarkably similar to these in the cuprates \cite{Zhou,Analytis}. The quantum critical fluctuations of one of the heavy fermion compounds, measured by neutron scattering  \cite{Schroeder,Schroeder2,Si} has also been fitted to a form of local-critical fluctuations \cite{footnote}. All these problems share the property that they are highly anisotropic so that the fluctuation problem may be regarded as two-dimensional in space. The microscopic physics in these problems is of course quite different. The universality class of local quantum criticality appears to encompass diverse physical systems.

The dissipative quantum XY model is a quantum generalization of the classical 2D XY model. The latter can be solved by integrating over the spin-wave variables to cast the model in terms of topological excitations, the vortices, which are responsible for the Kosterlitz-Thouless (KT) transition: vortices occur as bound pairs of zero net-vorticity at low temperatures while individual vortices proliferate in the high temperature disordered phase \cite{Berezinskii,Kosterlitz}.  In contrast, the dissipative quantum XY model has two additional features, (1) the kinetic energy of the fixed length 2D-rotors, and (2) dissipation of the (gradient of the) angular degrees of freedom. When dissipation is unimportant, the quantum model has been shown to have a quantum phase transition in the universality class of classical 3D XY model (with dynamical critical exponent $z=1$) \cite{Fisher} for the ratio of the kinetic energy parameter to the spatial coupling of the rotors above a critical value \cite{Janke90}. When dissipation is important, the problem may be usefully re-parameterized in terms of new degrees of freedom, which are two orthogonal sets of topological excitations, the vortices and the warps \cite{Aji2007} (see also Sec.~\ref{sec:anatrans}).  It is claimed that when the quantum phase transitions in the model are governed by proliferation of warps, the transitions are of the local-critical type and the critical fluctuations are of the form phenomenologically proposed for MFL~\cite{Varma1989}. It is important to have an unambiguous check of this solution of the model and its variants by other methods. The method used here is to simulate the model with the quantum Monte-Carlo method. 

In a set of Monte-Carlo calculations already done on the model \cite{Sudbo}, a rich phase diagram of the model was discovered. However, the correlations of the order parameter in some important regions were not studied. Nor were the properties of the model related to the topological excitations proposed \cite{Aji2007}. The results for all the quantities calculated here which were also calculated earlier \cite{Sudbo} are identical. In the present work, we calculate the correlation functions and relate them to those of the topological excitations which can be identified explicitly in the Monte-Carlo calculations.  We show that for transitions driven by warps, the order parameter fluctuations at the critical point are scale-invariant in imaginary time $\tau$, and calculate the behavior of the correlation functions as the critical point is approached. A result beyond those derived analytically is that the spatial correlations are consistent with a length scale $\xi_x$ which grows only logarithmically with the temporal length scale $\xi_{\tau}$. This is consistent with a dynamical exponent $z \to \infty$, making it a model in which local quantum-criticality is explicitly proven. 
 
We do not study, in this paper, the $z=1$ transition where the kinetic energy rather than the dissipation drives the transition, nor the passage between these two distinct types of transitions. That is clearly interesting and important but is reserved for future work.

This paper is organized as follows. We introduce the model and the details of quantum Monte-Carlo method in Sec. \ref{sec:modelmethod}. In Sec.~\ref{sec:anatrans}, we explain how we identify the warps and vortices in the calculations.  We  show the obtained phase diagram and summarize the properties of three distinct phases: the Disordered, Quasi-ordered, and Ordered phases, in Sec.~\ref{sec:phasediagram}. In Secs. \ref{sec:nortocsc}, \ref{sec:csctofsc}, and \ref{sec:nortofsc}, we show the calculated critical fluctuations of the order parameter at the transitions between them, and their relation to the change in density and correlations of warps and vortices across the transitions.   We focus especially on the transition from the disordered phase to the ordered phase (Sec.  \ref{sec:nortofsc}), and explicitly show that the fluctuations have a temporal correlation length exponentially larger than the spatial correlation length. In Sec. ~\ref{sec:h4}, we discuss the effect of a four-fold anisotropic field, relevant to cuprates and the anti-ferromagnets. Our conclusion and the directions for future analytical calculations are presented in Sec.~\ref{sec:conclusion}. 

\section{Model and Method}
\label{sec:modelmethod}

\subsection{(2+1)D Quantum dissipative XY model}
\label{sec:model}

The action of  the (2+1)D quantum dissipative- anisotropic XY model is ~\cite{Sudbo}
\begin{eqnarray}
S &=&-K_0 \sum_{\langle {\bf x, x}' \rangle} \int_0^{\beta} d \tau \cos(\theta_{{\bf x}, \tau} - \theta_{{\bf x}', \tau}) \nonumber \\
& +& \frac 1 {2E_c} \sum_{{\bf x}} \int_0^\beta d \tau \left( \frac{d \theta_{{\bf x}}}{d\tau}\right)^2  \nonumber \\
&+&  \frac{\alpha}{2} \sum_{\langle{\bf x, x}'\rangle} \int d \tau  d\tau' \frac {\pi^2}{\beta^2} \frac {\left[(\theta_{{\bf x}, \tau} - \theta_{{\bf x}', \tau})  -(\theta_{{\bf x}, \tau'} - \theta_{{\bf x}', \tau'}) \right]^2}{
\sin^2\left(\frac {\pi |\tau-\tau'|}{\beta}\right)} \nonumber \\
&-& h_4^0 \sum_{{\bf x}} \int d\tau \cos(4\theta_{{\bf x},\tau}),
\label{eq:model}
\end{eqnarray}
where ${\bf x}$ labels the coordinates of a lattice site in 2D spatial dimension and $\tau$ labels an imaginary time in extra temporal dimension. $\tau \in [0,\beta]$, where $\beta$ is the inverse of temperature $1/(k_B T)$. $\theta_{{\bf x},\tau}$ is the angle of the planar spin. $\langle {\bf x, x}'\rangle$ denotes nearest neighbors.  The first term is the spatial spin coupling term as in classical XY model. The second term is the kinetic energy where the charging energy $E_c$ serves as inertia. The third term describes quantum dissipations of the ohmic or Caldera-Leggett type~\cite{Caldeira1983}. The physical origin of such a term in the context of superconductor-insulator transitions \cite{chakra86, Fisher86}, and in the context of loop-current order in cuprates  \cite{Aji2007} has been discussed.  In the latter case, we note that the symmetry of the loop-current order is described by the Ashkin-Teller (AT) model with four discrete directions of the $\theta$-variables (or two Ising variables $\sigma^{x,y}$). We simulate the Ashkin-Teller model by imposing a strong four-fold anisotropic field $h_4^0$ (the fourth term).  We recall that the anisotropy is marginally irrelevant in the classical 2D-XY model \cite{Jose}. Strictly speaking, we should also add a term with interactions $\propto \cos[2(\theta_{{\bf x}, \tau} - \theta_{{\bf x}', \tau})]$ to represent the Ashkin-Teller model completely. Such a term has been shown to be irrelevant in the analytic calculations \cite{Aji2007} in the fluctuation regime. We have verified this assertion in the Monte-Carlo calculations. 

In numerical simulations, we choose a 2D square lattice with $N\times N$ lattice sites. Periodic boundary conditions are imposed along both $x$ and $y$ directions. We further discretize the imaginary time axis $[0,\beta]$ into $N_{\tau}$ slices. In this discretized (2+1)D lattice,  the action can be rewritten as 
\begin{eqnarray}
S &=& - K \sum_{\langle{\bf x},{\bf x}'\rangle, \tau} \cos(\Delta\theta_{{\bf x},{\bf x}',\tau}) 
+\frac {K_{\tau} } {2}  \sum_{\bf x, {\tau}}  ( \theta_{{\bf x},\tau }-\theta_{{\bf x},\tau-1})^2  \nonumber \\
&+& \frac{\alpha}2  \sum_{\langle{\bf x},{\bf x}'\rangle, \tau, \tau'} \frac {\pi^2}{ N_{\tau}^2} \frac {[\Delta\theta_{{\bf x},{\bf x}',\tau}-\Delta\theta_{{\bf x},{\bf x}',\tau'}]^2}{
\sin^2\left(\frac {\pi |\tau-\tau'|}{N_{\tau}}\right)} - h_4 \sum_{{\bf x},\tau} \cos (4\theta_{{\bf x},\tau} ),
\label{eq:modeld}
\end{eqnarray}
where $K_{\tau} \equiv 1/(E_C\Delta \tau)$, $K\equiv K_0\Delta \tau$,  $h_4=h_4^0\Delta\tau$, and $\Delta \tau = \beta/N_{\tau}$. We choose $K$, $K_{\tau}$, $\alpha$, and $h_4$ as independent dimensionless variables, and tune them separately. In other words, we consider the variables in units of the physical ultra-violet cut-off $\Delta \tau$ which is fixed. In this representation, the temperature is controlled by $N_{\tau}^{-1}$. The calculations are asymptotically correct for the quantum problem where $1/\beta = T \to 0$ or $N_{\tau} \to \infty$. This requires in practice that we ensure that the results converge in the range of $N_{\tau}$ actually studied. 
 
\subsection{Analytic transformation of the Model: Warps, Vortices etc.} 
\label{sec:anatrans}

It is useful to briefly review the analytic solution of the model in order to understand several aspects of the Monte-Carlo results including the physics of the three different phases found in Ref.~[\onlinecite{Sudbo}] and the mechanism for the (almost) spatial locality of the fluctuations.

In Ref. ~[\onlinecite{Aji2007}], it is shown that after making a Villain transformation \cite{Villain} and integrating over the small oscillations or spin-waves, the action is expressed in terms of link variables which are differences of $\theta$'s at nearest neighbor sites, as shown in Fig.~(\ref{fig:warp}).
\be
\label{eq:m}
m_{{\bf x,x}'}(\tau, \tau') \equiv \theta({\bf x},\tau) - \theta({\bf x}',\tau').
\ee
Further
\be
{\bf m} = {\bf m}_{\ell} +{\bf m}_t
\ee
where ${\bf m}_{\ell}$, is the longitudinal (or curl-free) part  and ${\bf m}_t$ is the transverse (or divergence-free) part . The appearance of ${\bf m}_{\ell}$ is a novel feature of the quantum dissipative XY-model.
Now define
\be
\nabla \times {\bf m}_t ({\bf x},\tau) = \rho_v({\bf x}, \tau) \hat{\bf z},
\ee
so that $\rho_v({\bf x}, \tau)$ is the charge of the vortex at $({\bf x},\tau)$, and
\be
\frac{\partial {\hat{\nabla}}\cdot {\bf m}_{\ell}({\bf x}, \tau)}{\partial \tau} = \rho_w({\bf x}, \tau).
\ee
$\rho_w({\bf x}, \tau)$ is called the ``warp'' at $({\bf x},\tau)$.

Although a continuum description is being used for simplicity of writing, it is important to do the calculation so that the discrete nature of the $\rho_v, \rho_w$  fields is always obeyed. In the numerical implementation of (2+1)D discrete lattice,  given the two bonds per site $({\bf x})$, one may construct a vector field ${\bf m}_{{\bf x},\tau}$, whose components are the two directed link variables in the Cartesian directions:  
\be
m^x_{i,j,\tau} &=& \theta_{i+1,j,\tau}-\theta_{i,j,\tau},  \nonumber \\
m^y_{i,j,\tau} &=& \theta_{i,j+1,\tau}-\theta_{i,j,\tau},  
\label{eq:vecfieldnum}
\ee
as shown in Fig.~(\ref{fig:warp}). Here ${\bf x}=(i,j)$.  

In terms of the vortex and warp densities, the action of the model was shown to be~\cite{Aji2007},
\be
\label{topomodel}
S &=& \sum_{{\bf k},\omega_n} 
\frac{K} {k^{2}}\left|\rho_{v} (\textbf{k},\omega_{n})\right|^{2}
-\frac{\alpha}
{4\pi\left|\omega\right|}\left|\rho_{w}(\textbf{k},\omega_{n})\right|^{2}
\\ \nonumber  &-& 
G\left(\textbf{k},\omega_{n}\right) \left( K K_{\tau}  - {\alpha
K_{\tau} \left|\omega_{n}\right|\over {4\pi }} - {\alpha^{2}k^{2}\over
{16\pi^{2}}}\right) 
\left|\rho_{w}(\textbf{k},\omega_{n})\right|^{2},
\ee
where 
\be
\label{G}
G\left(\textbf{k},\omega_{n}\right) =  {1\over{K k^{2} +
K_{\tau} \omega_{n}^{2}+\alpha\left|\omega_{n}\right|k^{2}}}.
\ee
The first term is the action of the {\it classical} vortices interacting with each other through logarithmic interactions in space but the interactions are local in time. The second term describes the warps interacting logarithmically in time but locally in space. In the third term, the terms proportional to $\alpha$ may be dropped in both the numerator and the denominator. Then this is just the action for a Coulomb field, which if present alone is known \cite{polyakov} not to cause a transition and is therefore  marginally irrelevant in the present problem. The warp and the vortex variables in the first two terms are orthogonal. With just these two terms alone, the problem is exactly soluble. If the first term dominates, one expects a transition of the class of the classical Kosterlitz-Thouless transition through binding of vortex-anti-vortex pairs in space but there is nothing to order the vortices with respect to each other in time. If the second term dominates, there is a quantum transition to a phase with binding of warp-antiwarp pairs in time but nothing to order them with respect to each other in space.  Given the ordering driven by either the density of isolated vortices or of isolated warps $\to 0$, the flow from one to the other is determined by the third term leading to possible ordering at $T=0$ both in time and space. It was derived in Ref.~[\onlinecite{Aji2007}] that when the ordering is driven through warps the fluctuations of the order parameter at the critical point have $1/\tau$ correlations, in time at the critical point [which on appropriate thermal Fourier transformation gives a spectral function $\tanh{(\omega/2T)}$].

\subsection{Quantum Monte-Carlo simulations} 
\label{sec:qmc}

We follow the numerical procedure as in Ref.~[\onlinecite{Sudbo}] for the Monte-Carlo simulations. To speed up the simulation, we choose $\theta_{{\bf x},\tau}$ to be a discrete variable, $n 2\pi/ 32$ ($n$ an integer), rather than a continuous variable. Adding more states does not affect the results, as found in Ref.~\onlinecite{Sudbo} and confirmed in our calculation.  The system size typically chosen is $N=50$ and $N_{\tau}=200$, which are found to be adequate for the parameter ranges not too close to the critical points. Other system sizes are also used in scaling analysis calculations.  

We start from a random configuration of $\{\theta_{{\bf x},\tau}\}$. To update the configuration, we sequentially sweep the lattice sites to update locally $\theta_{{\bf x},\tau}$ to $\theta_{{\bf x},\tau}+ \theta'$, where $\theta'$ is a random angle between $-2\pi$ and $2\pi$. We make measurements of the physical quantities of interest after 10 sweeps. We also employ parallel tempering technique to speed up the relaxation. The acceptance rate for this local update ranges from 46\%(disordered state) to 16\%( ordered state) in the range of parameters being calculated. We typically choose O($10^4$) warm up sweeps and O($10^6$) measurements in our Monte-Carlo simulations. For large enough measurements, the desired thermodynamic averages and correlation functions are well approximated. 

The following quantities are calculated to characterize the different phases and the transitions between them. 

{\it Action susceptibility.} The action susceptibility is defined as 
\begin{equation}
\chi_S = {1\over N^2 N_{\tau} } \left(\left\langle S^2\right\rangle -\left\langle S \right\rangle ^2\right),  
\end{equation} 
where $\langle \ldots \rangle$ denotes averaging over the $O(10^6)$ Monte-Carlo measurements. 
In classical systems, as $S=\beta H$, $\chi_S$ is related to the specific heat, $\chi_S = C_V/k_B$.  At $T \to 0$, it is a measure of zero-point fluctuations which are expected to be singular at the critical point due to the degeneracy in the spectra. 

{\it Helicity Modulus.}  The helicity modulus or spatial stiffness is defined from the change of energy resulting from the slow twist of spins along the spatial direction, or
\begin{eqnarray}
\Upsilon_x &=& {1 \over N^2 N_{\tau}} \left\langle \sum_{\langle{\bf x},{\bf x}'\rangle} \sum_{\tau} \cos(\Delta \theta_{{\bf x},{\bf x}', \tau}) \right\rangle  
  \nonumber \\
  &-& {K \over N^2 N_{\tau}} \left\langle \left( \sum_{\langle{\bf x},{\bf x}'\rangle} \sum_{\tau} \sin(\Delta \theta_{{\bf x},{\bf x}', \tau}) \right)^2 \right\rangle .
\end{eqnarray}
In the disordered state, the two terms have comparable contributions and $\Upsilon_x \to 0$. In an ordered phase, the second term vanishes while $\Upsilon_x$ becomes finite. 

{\it Order parameter.} For XY spins, the order parameter ${\bf M} ({\bf x},\tau) = (\cos \theta_{{\bf x},\tau}, \sin \theta_{{\bf x},\tau})$. Its modulus, the magnetization in the plane, is defined as 
\begin{equation} 
M = \frac {1}{N^2N_{\tau}} \left\langle \left| \sum_{{\bf x},\tau}  e^{i \theta_{{\bf x}, \tau}} \right| \right\rangle.
\end{equation}
In classical 2D XY model, the ordered phase has a quasi long-range order, where $M \sim (1/N)^{1/(8\pi K)}$, vanishes for $N \to \infty$. A question which we will be able to answer is whether there is a finite magnetization in the infinite size limit for the quantum  dissipative XY model.  We also found it illuminating to calculate $M_{2D}$, the magnitude of magnetization in the planes at a given time $\tau$ and then average it over the $\tau$. This is equivalent to finding the Kosterlitz-Thouless order parameter at each time slice and then averaging over the time slices. 
\begin{equation} 
M_{2D} = \frac {1}{N^2N_{\tau}} \left\langle \sum_{\tau} \left| \sum_{{\bf x}}  e^{i \theta_{{\bf x}, \tau}} \right| \right\rangle.
\end{equation}
By definition $M \leqslant M_{2D}$. Also $M=M_{2D} \ne 0$ (for $N \to \infty$) only if there is perfect long range order across time as well as space. 

{\it Correlation Function of the Order Parameter.} The principal results for the quantum-critical fluctuations are given by the order parameter correlation functions:
\begin{equation}
G_{\theta} ({\bf x}, \tau) = \frac {1}{N^2N_{\tau}} \sum_{{\bf x}',\tau'}\left\langle e^{i (\theta_{{\bf x'}+{\bf x}, \tau'+\tau} - \theta_{{\bf x}', \tau'})}\right\rangle.
\end{equation}
$G_{\theta} ({\bf x} \to \infty, \tau \to \infty) \to M^2$ while $G_{\theta} ({\bf x} \to \infty, \tau =0) \to M_{2D}^2$.  In Ref. [\onlinecite{Sudbo}], mean-square displacements in time $W^2_{\Delta \theta}$ are shown, which we have reproduced. These can also be obtained from the second moment of the above correlation function at ${\bf x} =0$. 

\begin{figure*}
\centering
\includegraphics[width=0.8\textwidth]{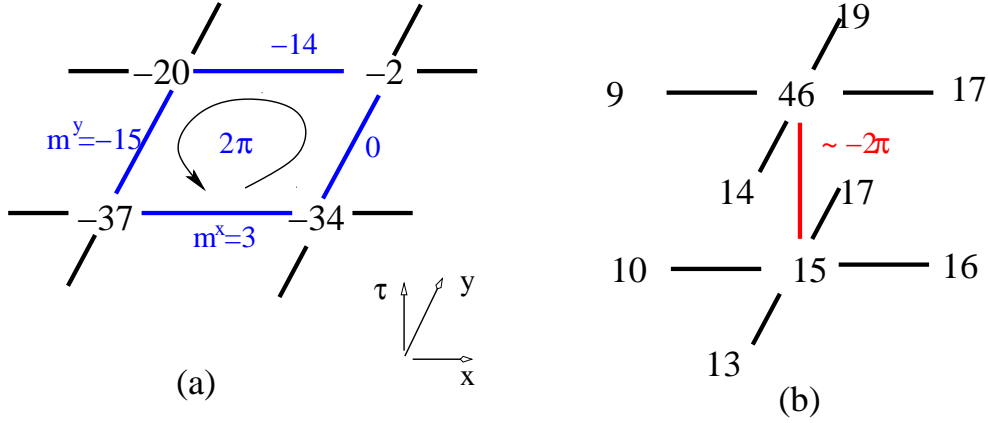}
\caption{Examples of vortex (a) and warp (b) excitations in numeric simulation. The numbers at the lattice points (in space or time) are the $\theta$'s in units of $2\pi/32$ and are non-compact variables. The numbers in the links are the velocity fields, i.e. the difference of $\theta$'s that a link connects. (a) For the plaquette shown, $(\nabla \times {\bf m})_{i,j,\tau}$ is 32, or $2\pi$, showing a vortex. In (b), the change of $({\hat \nabla}\cdot {\bf m})_{i,j,\tau}$ for two neighboring time slices is close to $-2\pi$, showing an antiwarp.  
}
\label{fig:warp}
\end{figure*}

{\it  Vortices and warps: densities and self/mutual correlations.} The curl of the vector field ${\bf m}$  can be calculated numerically from the four link variables of a plaquette,   
\be
\label{vortex}
\rho_v({\bf x},\tau) = \frac{1}{2\pi}\left(\nabla \times {\bf m}\right)_{i,j,\tau} = \left(m^x_{i,j, \tau} + m^y_{i+1,j, \tau} - m^x_{i+1,j+1, \tau}-m^y_{i,j+1, \tau}\right)/(2\pi),
\ee
where we restrict $m^{x,y}$ to be within $(-\pi, \pi)$ by adding or subtracting $n2\pi$. If $(\nabla \times {\bf m})_{i,j,\tau}=\pm 2\pi$, we identify a vortex/antivortex, or $\rho_v({\bf x}, \tau)=\pm 1$. Similarly, the divergence of the vector field can be calculated from four links connected to the site
\be  
(\hat{\nabla}\cdot {\bf m})_{i,j,\tau}&=& (m^x_{i,j, \tau} - m^x_{i-1,j, \tau} + m^y_{i,j, \tau}-m^y_{i,j-1, \tau})/4.  
\ee
We therefore use the following criterion to identify a warp (antiwarp) charge 
\begin{eqnarray}
\label{warp}
\rho_w({\bf x},\tau)&=&1, \text { if }({\hat \nabla}\cdot {\bf m})_{i,j,\tau+1}- ({\hat \nabla}\cdot {\bf m})_{i,j,\tau} > 2\pi -\delta\theta \nonumber \\
\rho_w({\bf x},\tau)&=&-1,  \text{ if } ({\hat \nabla}\cdot {\bf m})_{i,j,\tau+1}- ({\hat \nabla}\cdot {\bf m})_{i,j,\tau} < -2\pi +\delta\theta. 
\end{eqnarray}
where $\delta\theta \ll 2\pi$ to accommodate small angle changes due to spin waves. Examples of vortices and warps are also shown in Fig.~(\ref{fig:warp}). 

After identifying the vortex and warp charges $\rho_{v,w}({\bf x}, \tau)$ in the system,  we can calculate their densities, 
\be
\rho_{v,w} = \frac{1}{N^2N_{\tau}}\sum_{{\bf x},\tau} \left\langle| \rho_{v,w}({\bf x}, \tau)|\right\rangle,
\ee
as well as  their correlation functions:  
\begin{eqnarray}
G_{v,w} ({\bf x}, \tau) = \frac{1}{N^2N_{\tau}} \sum_{{\bf x}',\tau'} \langle \rho_{v,w}({\bf x}'+{\bf x} , \tau'+\tau) \rho_{v,w}({\bf x}', \tau') \rangle . 
\end{eqnarray}

Charge neutrality for both vortices and warps should be preserved. We verify this by calculating the net density $\delta \rho_{v,w} = [\sum_{{\bf x},\tau} \langle \rho_{v,w}({\bf x}, \tau)\rangle]/(N^2N_{\tau})$, and find that in practice, $|\delta \rho_v| /\rho_v < 10^{-5}$ and $|\delta \rho_w |/\rho_w < 10^{-2}$. 
To capture the correlations between warps and vortices, we also calculate
\begin{eqnarray}
G_{vw} &=& \frac{1}{N^2N_{\tau}} \sum_{{\bf x},\tau} \langle |\rho_v({\bf x}, \tau) \rho_w({\bf x}, \tau)|\rangle,
\end{eqnarray}
i.e., the probability to find vortices in the vicinity of a warp and vice versa. If warps and vortices are not correlated, we expect $G_{vw} =\rho_v\rho_w$. 

\section{Summary of the Phase Diagram}
\label{sec:phasediagram}

 \begin{figure*}[tbh]
\centering
\includegraphics[width=0.8\textwidth]{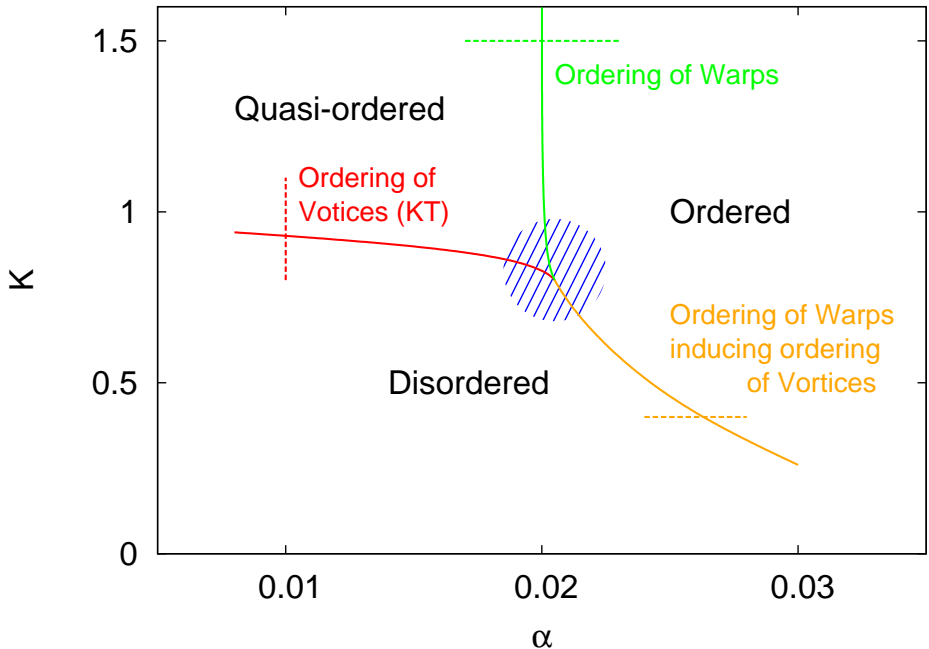}
\caption{Phase diagram for the quantum dissipative XY model in $\alpha-K$ plane. Here $K_{\tau}=0.01$. The transition points are determined from the non-analyticity in various static properties with a system size $N=50$ and $N_{\tau}=200$ (the transition points for infinite systems can be determined by a finite-size analysis). The area where the lines join (blue shaded area) has not been explored thoroughly enough to precisely determine how the phase boundaries meet. The dashed lines show the sweep of parameters presented in the following sections for the study of  correlations across the three types of transitions. The phase diagram is obtained by several such sweeps across each transition.}
\label{fig:pd}
\end{figure*}

\begin{figure*}[tbh]
\centering
\includegraphics[width=0.7\textwidth]{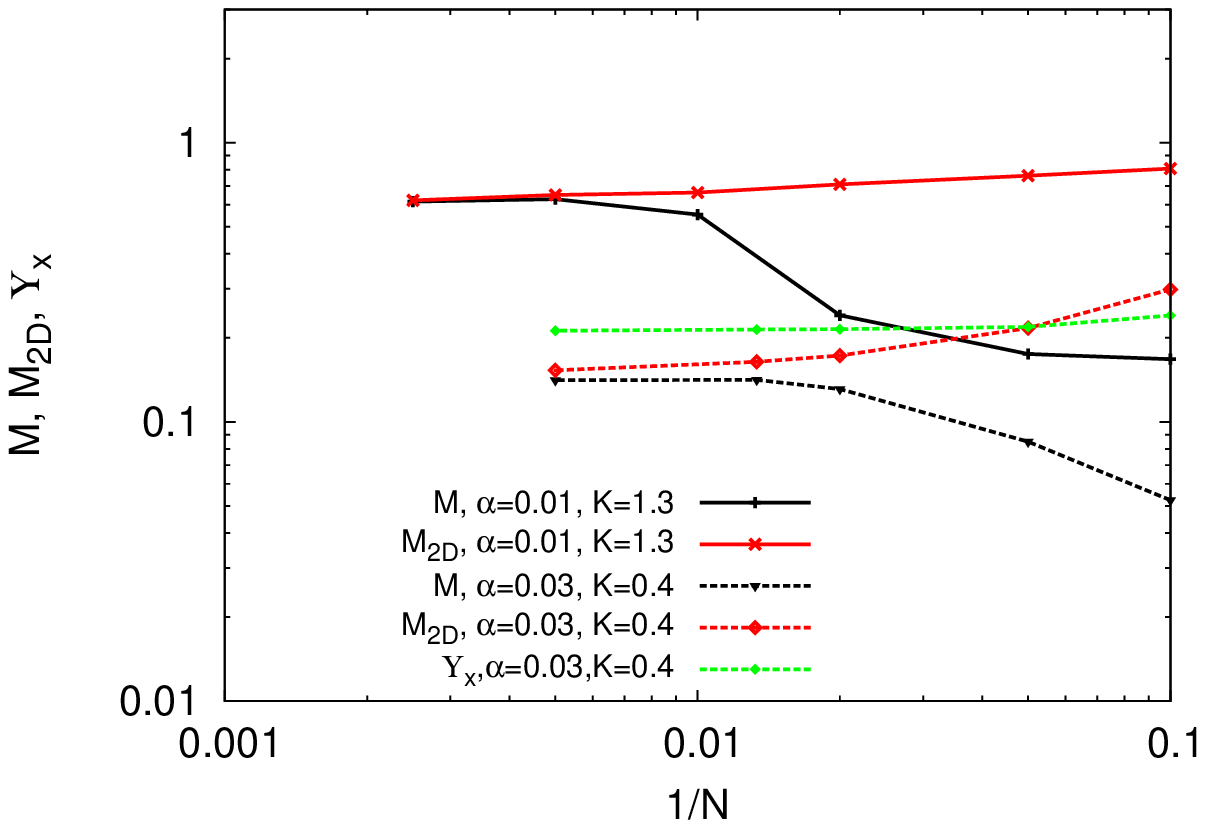}
\caption{The spatial size dependence of $M_{2D}$ and $M$ in the Quasi-ordered (top two curves) and Ordered phases (bottom 2 curves). Also shown is $\Upsilon_x$ in the Ordered phase.  The parameters chosen for the Quasi-ordered phase are $K_{\tau}=0.01$, $\alpha=0.01$, $K=1.3$ and $N_{\tau}=20$, while for the Ordered phase, $K_{\tau}=0.01$, $\alpha=0.03$, $K=0.4$ and $N_{\tau}=100$. The slow decrease of the top most curve ($M_{2D}$) is just the finite size scaling to 0 at $N \to \infty$  in the Kosterlitz-Thouless phase \cite{Bramwell1994}, which $M$ asymptotically joins. Their behaviors are different in the Ordered Phase, where asymptotically, they are both consistent with a finite value at $N \to \infty$. The difference in $\Upsilon_x$ across the Disordered to Quasi-Ordered and across the Disordered to Ordered phases is discussed in the text. }
\label{fig:mmssize}
\end{figure*}

We first study the dissipative quantum XY model [cf. Eq.~(\ref{eq:modeld})] without the four-fold anisotropic field $h_4$, whose effect is addressed in Sec.~\ref{sec:h4}.  We focus on the transitions driven by dissipations, for which a small kinetic energy parameter $K_{\tau}$ is chosen. The phase diagram in $\alpha$-$K$ plane with fixed $K_{\tau}=0.01$  is given in Fig.~(\ref{fig:pd}). It is similar to the $K_{\tau}=0.002$ phase diagram obtained in Ref.~[\onlinecite{Sudbo}]. Here, three distinct phases are identified, a ``Disordered" phase,  a ``Quasi-ordered" phase, and an ``Ordered" phase (named as NOR, CSC, and FSC phases, respectively, in Ref.~[\onlinecite{Sudbo}]). Their properties are summarized in Table \ref{tab:threephases}. The Disordered phase has short-ranged correlations in both the spatial and temporal directions. The Quasi-ordered phase, while also have short-ranged temporal correlations, has a quasi long range order in 2D spatial plane (for each time slice), consistent with KT spatial order. $M_{2D}$ is finite and falls off slowly for large $N$, as shown in Fig.~(\ref{fig:mmssize}). The order parameter $M$ follows $M_{2D}$ asymptotically for $N\gg N_{\tau}$ while $M\to 0$ for $N \ll N_{\tau}$.  The Ordered phase has long range order in both spatial and temporal directions, where  $M$ goes to a finite value as $N, N_{\tau}  \to \infty$. 

 The transition from the Disordered phase to the Quasi-ordered phase can be achieved by increasing $K$ at small $\alpha$. As the temporal correlations remain relatively unchanged across the transition, the transition is characterized by the spatial ordering as in KT transition, due to binding of  vortex of anti-vortex pairs. For increasing $\alpha$, the system also orders in time, leading to a transition from the Quasi-ordered phase to the Ordered phase. For small $K$, there is a direct phase transition from the Disordered to the Ordered phase. This is in general in accord with the discussion in the previous paragraph based on the properties expected for the topological model of Eq.~(\ref{topomodel}). We will show that the transition from the Quasi-ordered to the Ordered phase (in Sec. \ref{sec:csctofsc}) as well as that from the Disordered phase to the Ordered phase (in Sec. \ref{sec:nortofsc}) occur primarily through freezing of warps. In the second transition, the vortices freeze as an accompaniment to the freezing of warps, in a manner distinct from that at the KT transition. 

\begin{table}[tbh]
\begin{tabular}{c|c|c|c}
\hline
Quantity & Disordered & Quasi-ordered & Ordered \\
\hline \hline
$M$ &  $0$ & decreases $\to 0$ for $N \to \infty$   & finite \\
$\rho_v$ & O(1) & $\ll$ 1 & $\ll$1 \\
$G_{v} (x)$ & exponential & power law & power law\\
$\Upsilon_x$ & 0 & finite, jump at transition & finite, no jump at transition \\
$\rho_w$ & O(1) & O(1) & $\ll$ 1 \\
$G_{w} (\tau)$ & $1/\tau^2$ & $1/\tau^2$ & $1/\tau^\alpha (\alpha > 2 )$ \\
$G_{\theta} (x, 0)$ & exponential & quasi-long range & long range \\ 
$G_{\theta} (0, \tau)$ & exponential & exponential & long range \\
\hline 
\end{tabular}
\caption{Characteristic properties of three phases of the dissipative XY model. Definitions of these quantities are provided in Sec. \ref{sec:qmc}. }
\label{tab:threephases}
\end{table}

\section{Transition from the Disordered phase to the Quasi-Ordered phase}
\label{sec:nortocsc}

\begin{figure*}[tbh]
\centering
\includegraphics[width=0.55\textwidth]{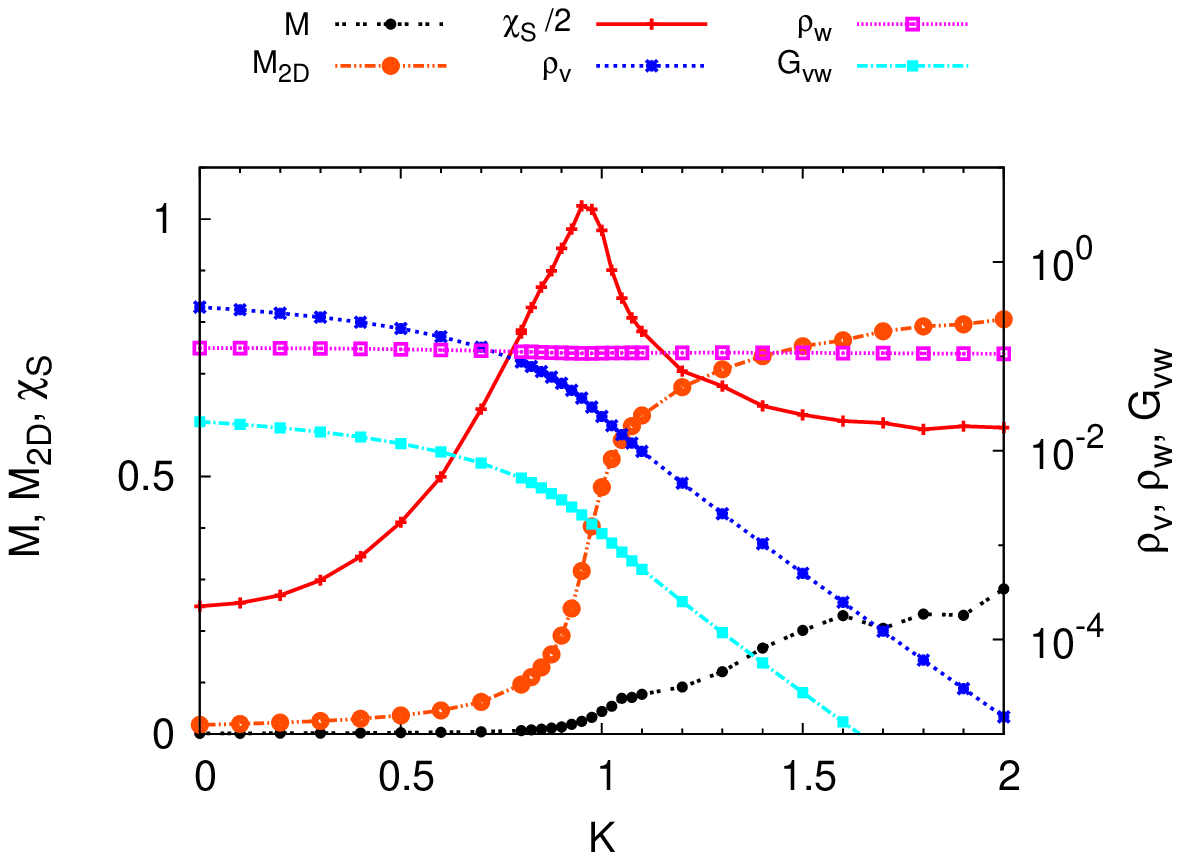}
\includegraphics[width=.45\textwidth]{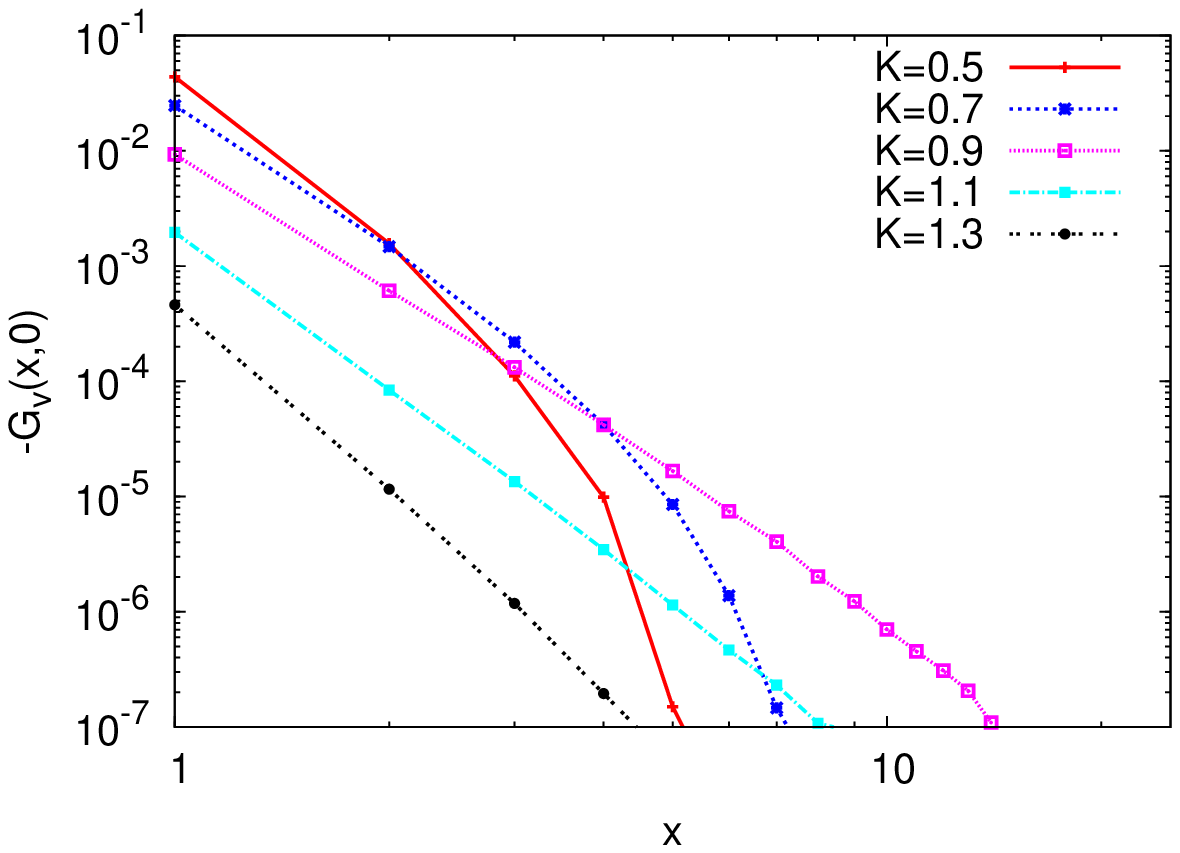}
\includegraphics[width=.45\textwidth]{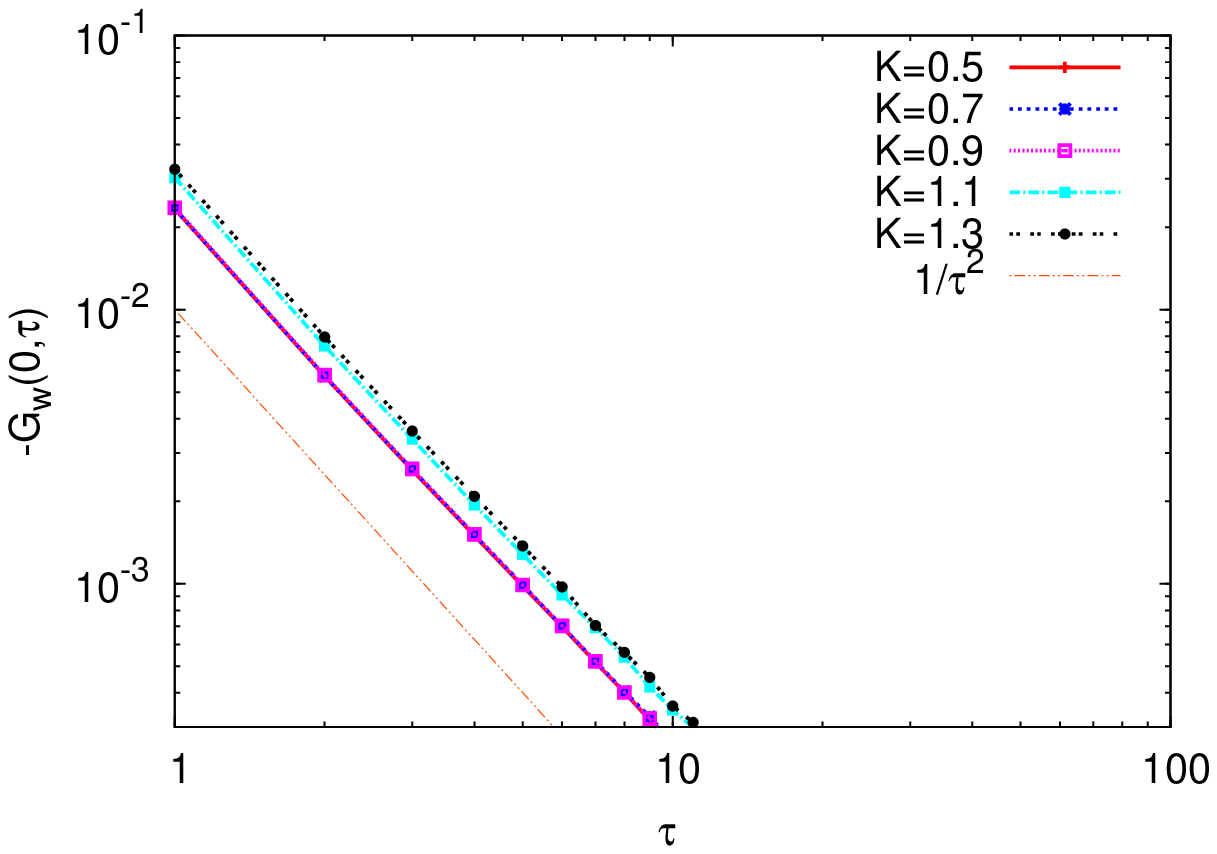}
\caption{Static properties (top panel), vortex and warp correlation functions (bottom panels) of transition from the Disordered phase to the Quasi-ordered phase. Here, $K_{\tau}=0.01$, $\alpha=0.01$, and $K$ is varied. The results shown are for $N=50$ and $N_{\tau}=200$. Note that some quantities are scaled to fit in the figure. The vortex density changes rapidly below the transition while the warp density remains smooth. Other aspects of the transition and of the Quasi-ordered phase are discussed in the text.}
\label{fig:12static}
\end{figure*}

The transition from the Disordered phase to the Quasi-ordered phase is studied by fixing $\alpha=0.01$ and varying $K$.  The static properties are shown in Fig.~(\ref{fig:12static}). We find that above a critical value  $K_c$, which weakly depends on $\alpha$, the spatial magnetization $M_{2D}$ becomes finite.  As shown in Fig. ~(\ref{fig:mmssize}), $M_{2D}$ decreases slowly when $N$ increases. As discussed later, this decrease is consistent with the logarithmic decrease found in earlier calculations \cite{Bramwell1994}. $M \to M_{2D}$ when $N_{\tau} \ll N$ and $M \to 0$ when $N_{\tau} \gg N$. The difference between $M$ and $M_{2D}$ is also reflected in the order parameter correlations in the time direction, which shows oscillatory features at long times (not shown). This phase has only quasi long-range (power law) spatial order. As shown in Ref.~[\onlinecite{Sudbo}], the helicity modulus $\Upsilon_x$ becomes finite in the Quasi-ordered phase. Finite size scaling of the helicity modulus $\Upsilon_x$ shows a Nelson-Kosterlitz \cite{Nelson-Kosterlitz}  jump at $K_c$. This is related to the vortex density seen in Fig.~(\ref{fig:12static}) , which decreases with increasing $K$,  and changes slope at $K_c$. These are consistent with KT transition in classical 2D XY model. Meanwhile, we find that in the temporal direction, all quantities remain relatively unchanged from those in the disordered phase.  The vortex-warp correlation $G_{vw} \approx \rho_v\rho_w$, indicating vortices and warps are not correlated, in either the Disordered phase or the Quasi-ordered phase. 

We also plot the correlation functions of warps and vortices in Fig.~(\ref{fig:12static}). For the equal-time vortex correlation  $G_v(x,0)\equiv G_v(x, y=0, \tau=0)$, $G_v(0,0) = \rho_v > 0$ (not shown due to the logarithmic scale) while $G_v(x\neq0,0) <0$, reflecting that the vortex-antivortex correlations dominate at long distance. When $K$ increases, $-G_v(x)$ changes from an exponential decay in the disordered phase to a power-law decay in Quasi-Ordered phase. These are consistent with the KT transition as well. The warp correlation along temporal direction at the spatial site $G_w(0,\tau)$ also satisfies $G_w(0,0)=\rho_w > 0$ and $G_w(0,\tau \neq0) <0$.  In this transition, it remains unchanged in asymptotic form $\propto 1/\tau^2$.  

\begin{figure*}[tbh]
\centering
\includegraphics[width=0.55\textwidth]{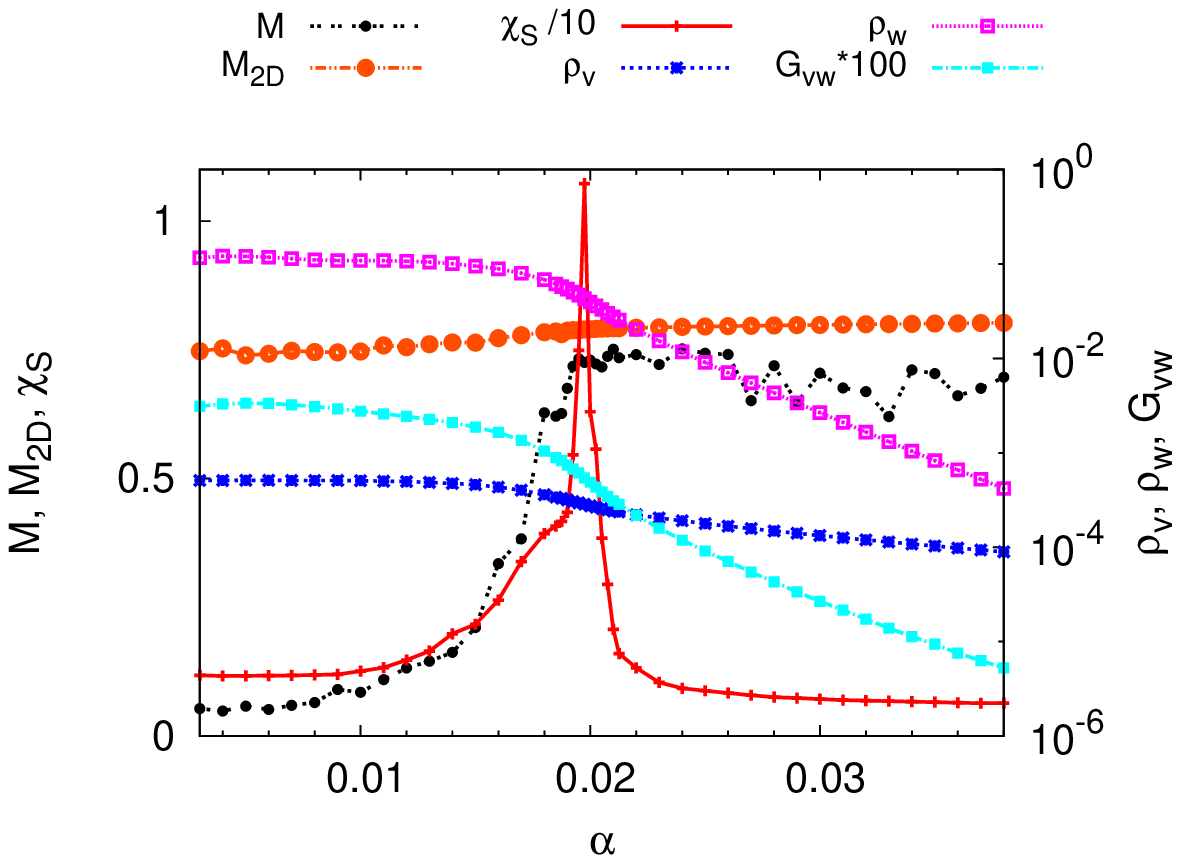}
\includegraphics[width=.45\textwidth]{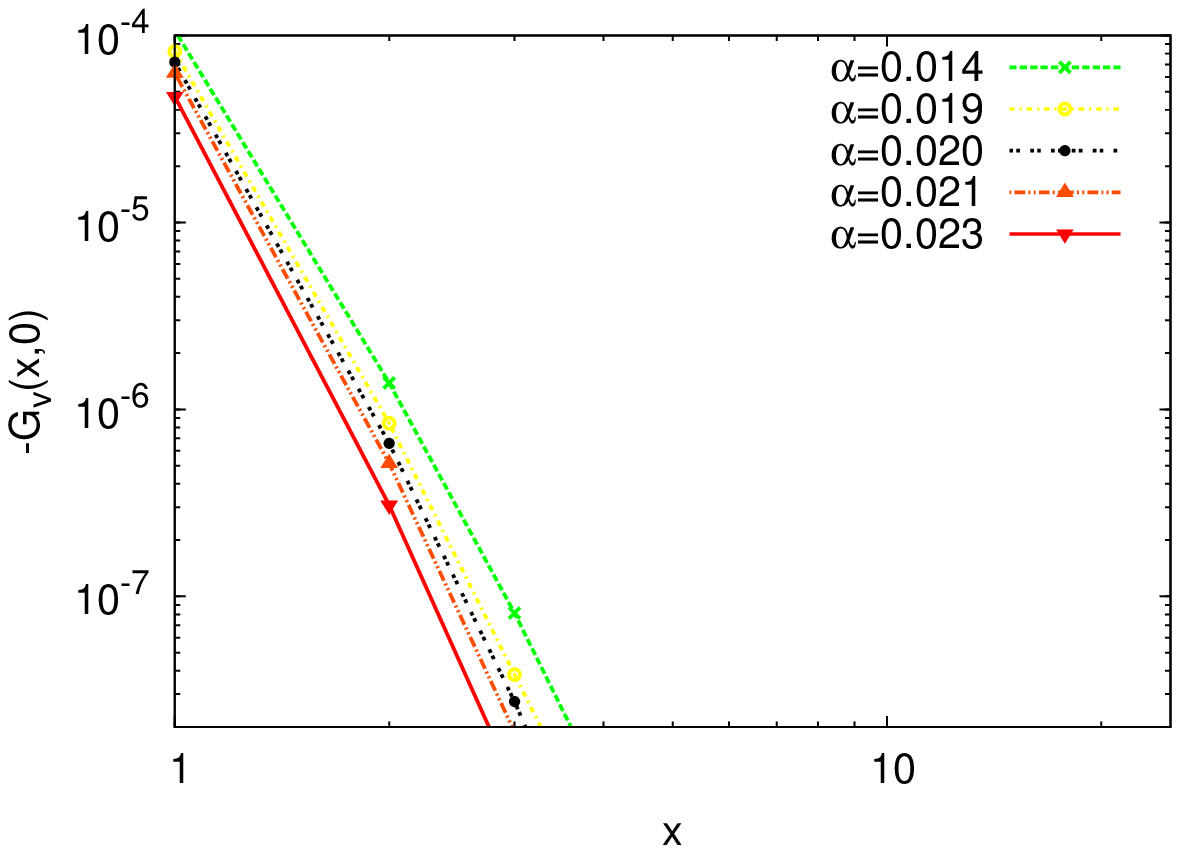}
\includegraphics[width=0.45\textwidth]{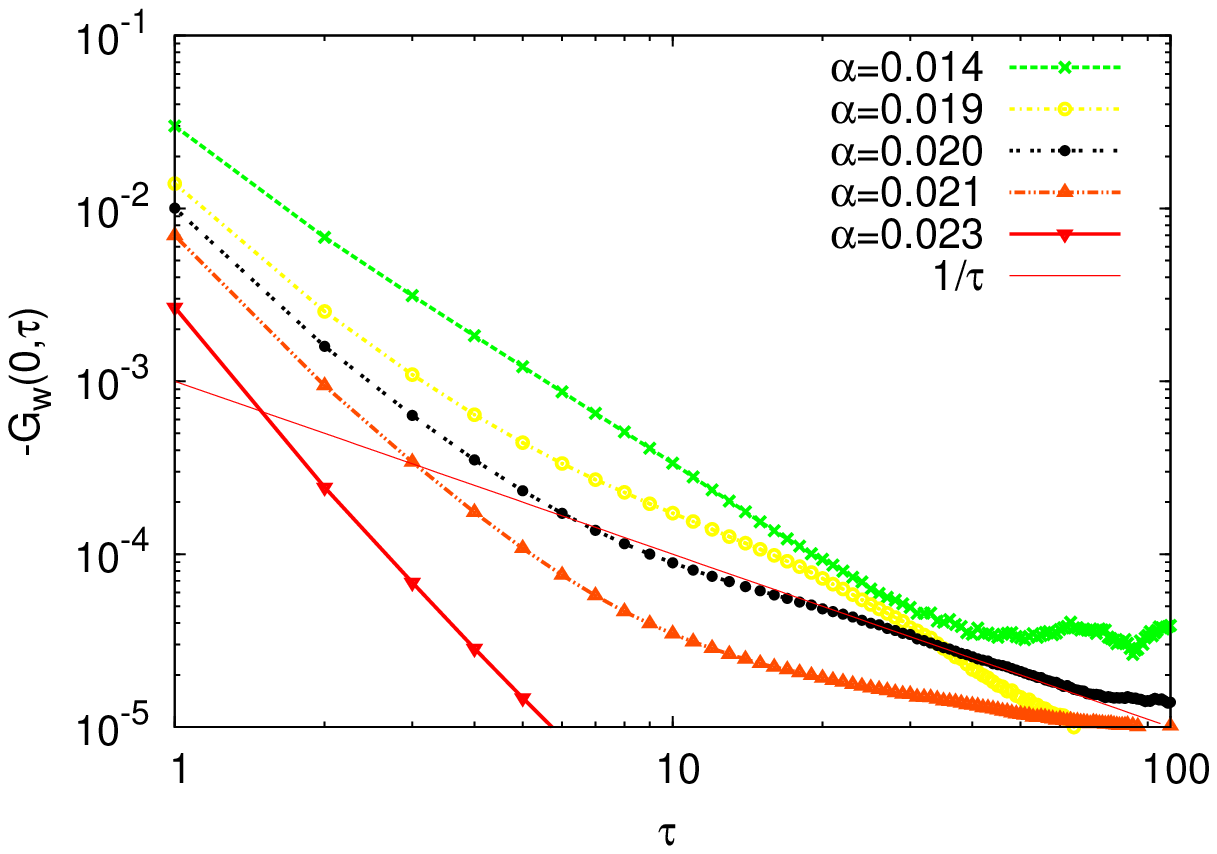}
\caption{Static properties (top panel), vortex and warp correlation functions (bottom panels) of transition from the Quasi-ordered phase to the Ordered phase. Here, $K_{\tau}=0.01$, $K=1.5$, and $\alpha$ is varied. The results shown are for $N=50$ and $N_{\tau}=200$.  }
\label{fig:23static}
\end{figure*}

\section{Transition from the Quasi-Ordered phase to the Ordered phase}
\label{sec:csctofsc}

We choose a suitable $K$ and tune the transition from the Quasi-Ordered phase to the Ordered phase by increasing the dissipation strength $\alpha$. Various static properties as functions of $\alpha$ and  correlation functions for selected $\alpha$'s are shown in Fig.~(\ref{fig:23static}).  The peak in the action susceptibility $\chi_S$ implies a phase transition at $\alpha_c \approx 0.02$. We find that properties characterizing spatial orders, such as $M_{2D}$, $\rho_v$ and $\Upsilon_x$, have small non-analytic changes, as already discovered in Ref. ~[\onlinecite{Sudbo}]. The significant changes are properties characterizing temporal order.  The asymptotic behavior of the warp density  is similar to that of vortex density at KT transition: it changes slope at $\alpha_c$ and decreases exponentially as $\alpha$ further increases.  $M$ keeps increasing and saturates to $M_{2D}$ at $\alpha \gg \alpha_c$ (at large system sizes). The warp correlation functions decay faster for larger $\alpha$, changing from $1/\tau^2$ in the Quasi-Ordered phase to $1/\tau^{a}$ ($a \sim 3$ for $\alpha=0.023$ in the figure) in the Ordered phase.  This indicates that warps and anti-warps, which are free in the Quasi-Ordered phase, also are bound in the ordered phase. Near $\alpha_c$, a slower decay at large times is observed. As shown in the figure, it can be fitted as $1/\tau$. This is in agreement with the analytical analysis \cite{Aji2007}. While the vortex-warp correlation $G_{vw} \sim \rho_v \rho_w$ in the Quasi-ordered phase, we find $G_{vw} > \rho_v \rho_w$ in the Ordered phase, and their difference increases when $\alpha$ is further increased from $\alpha_c$.  This implies that vortices and warps are correlated inside the Ordered phase. 

\section{Transition from the Disordered phase  to the Ordered phase}
\label{sec:nortofsc}

This is the part of the problem which we shall discuss most thoroughly. We show results for  a suitable value, $K=0.4$ and tune $\alpha$ across the transition at $\alpha = \alpha_c(K)$. Similar results have been obtained for other values of these parameters across the transition, keeping $K_{\tau} = 0.01$ fixed at this low value. Note the fine scale on which $\alpha$ is varied compared to $K$ in Fig. (\ref{fig:pd}) to tune across the transition. 

\subsection{Static properties and correlations}
\label{sec:nortofscprop}

\begin{figure*}[tbh]
\centering
\includegraphics[width=0.55\textwidth]{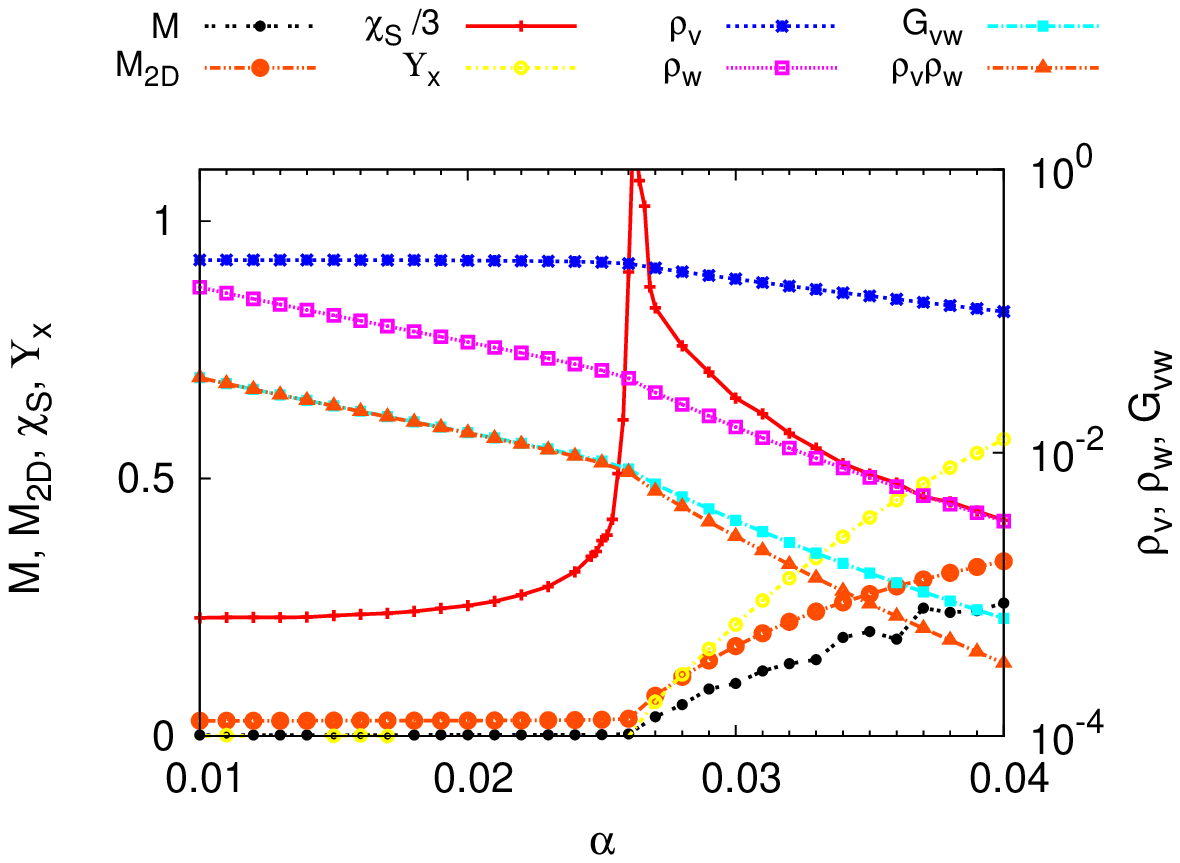}
\includegraphics[width=0.45\textwidth]{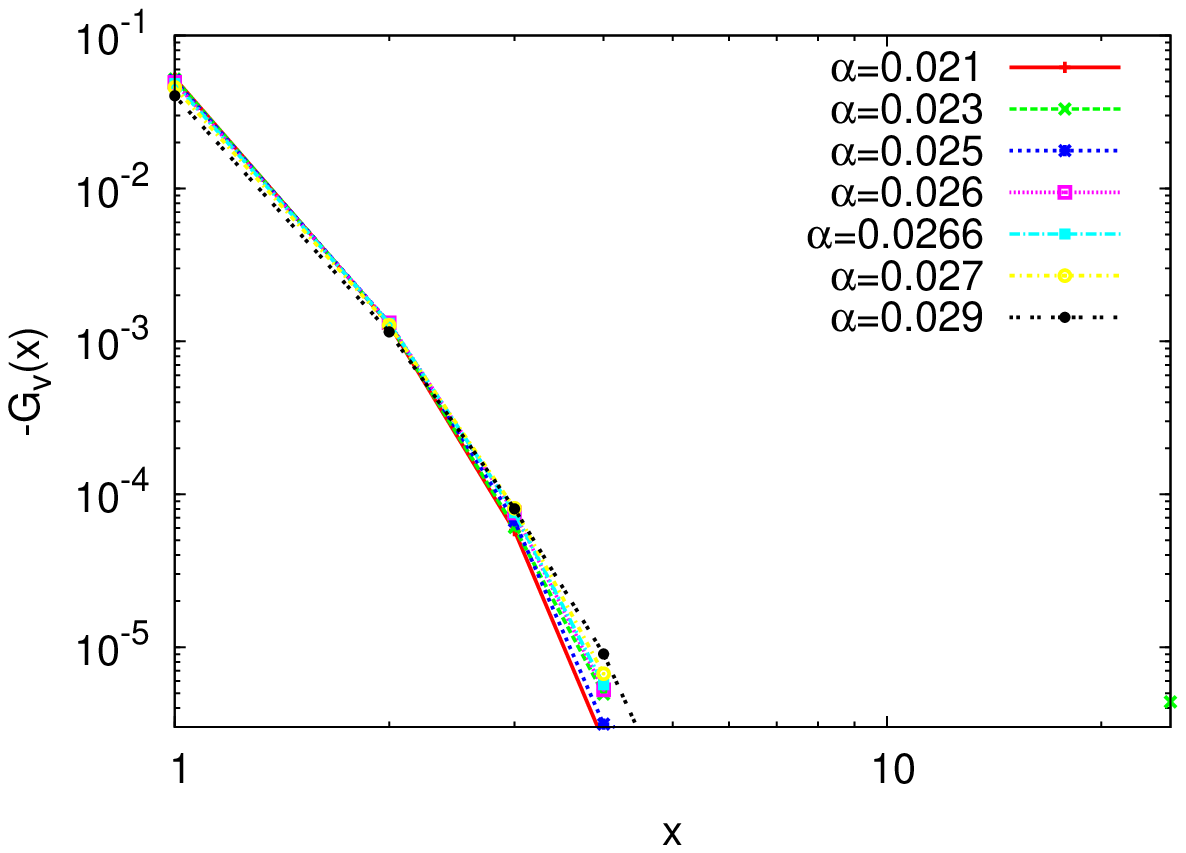}
\includegraphics[width=0.45\textwidth]{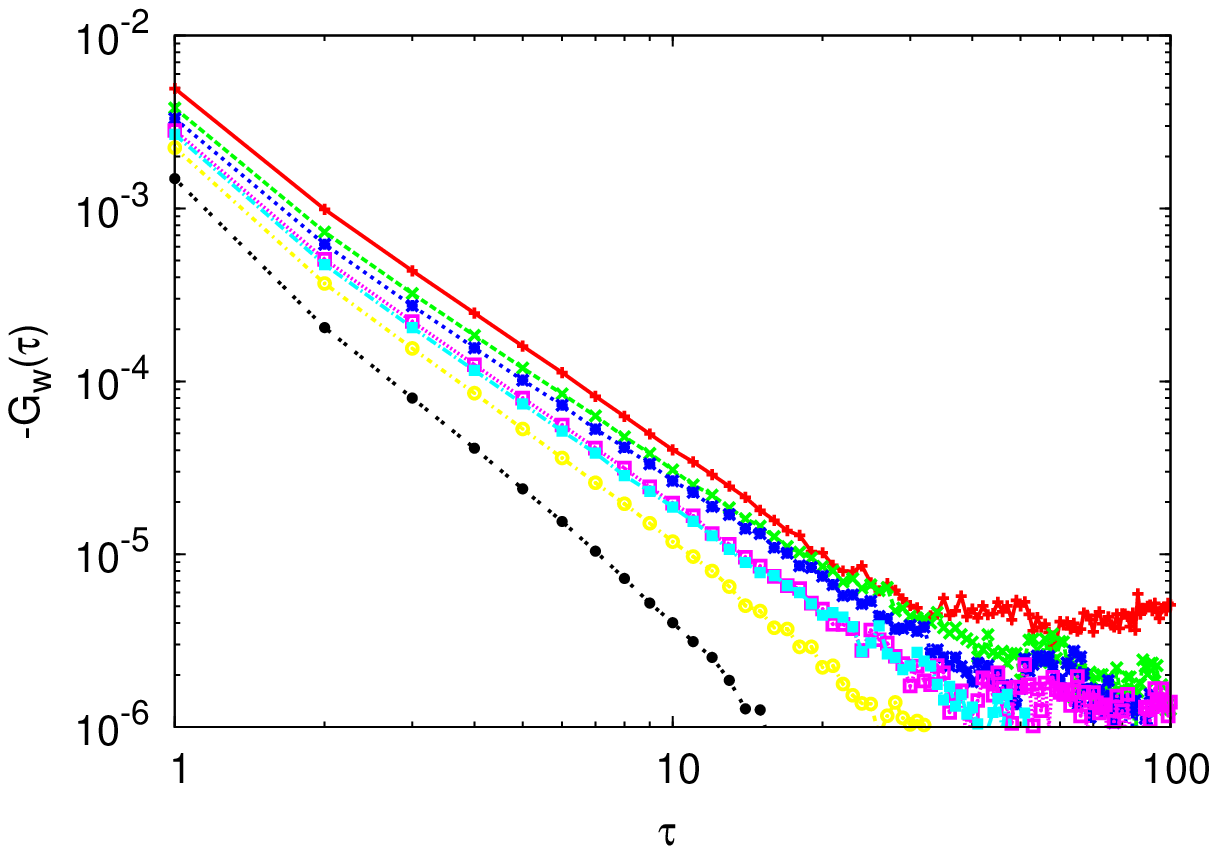}
\caption{Static properties (top panel),  vortex and warp correlation functions (bottom panels) of transition from the Disordered phase to the Ordered phase. Here, $K_{\tau}=0.01$, $K=0.4$, and $\alpha$ is varied. The results shown are for $N=50$ and $N_{\tau}=200$. We find that when $M_{2D}$ becomes finite, $M$ varies with the size and only approaches $M_{2D}$ at large system size [see Fig.~(\ref{fig:mmssize})]. The vortex and warp correlations are discussed in the text.}
\label{fig:13static}
\end{figure*}

The static properties shown in Fig.~(\ref{fig:13static}) are all non-analytic near $\alpha_c \approx 0.0260$. We estimate an uncertainty of $\pm 0.0002$ in $\alpha_c$, due to finite size effects.  The helicity modulus $\Upsilon_x$ and magnetization $m$ become finite for $\alpha > \alpha_c$. We notice that both the vortex and the warp densities change slope across $\alpha_c$. So long-range order appears to develop simultaneously along both the spatial and the temporal directions. However, on the disordered side, the warp density decreases by an order of magnitude as the transition is approached while the vortex density remains unchanged. This indicates a large critical region in which the temporal correlations are expected to grow while the spatial correlations remain short-range. On the ordered side, for $\alpha > \alpha_c$, the warp density has a more rapid change than the vortex density.  We also plot $\rho_v\rho_w$ explicitly to be compared with the mutual correlation between vortices and warps $G_{vw}$. 
We observe that $G_{vw} \propto \rho_w$ and $G_{vw}> \rho_v\rho_w$ when $\alpha>\alpha_c$, i.e., suggesting coupling of vortices to warps inside the ordered phase, while their difference $G_{vw}-\rho_v\rho_w$ vanishes at the critical point and becomes invisible on the disordered side. The study of correlation functions below will show that the spatial correlations do develop on the disordered side but with an exponentially slower dependence on $(\alpha_c-\alpha)$ than the temporal dependences. These facts suggest that the transition is driven by the quantum-freezing of warps. We speculate that this occurs through the third term in the action (\ref{topomodel}), which drives the fugacity of the vortices so that they also freeze. 

The self correlation functions of vortices and warps are also shown in Fig.~(\ref{fig:13static}).  The vortex correlation functions are relatively unchanged as $\alpha$ changes across the transition compared to the warp correlation functions, which have similar changes as in the Quasi-ordered phase to the Ordered phase transition.  Near $\alpha_c$, the latter shows slower decay at long times. However, whether it has the $1/\tau$ behavior requires a calculation with larger time slices and more iterations to demonstrate. 

\subsection{Size dependence of $\Upsilon_x$ and $M$}
\label{sec:size}

We have also studied the difference of the vortex freezing across the Disordered to Ordered phase transition compared to that across the Disordered to Quasi-ordered phase transition(which is of the pure KT type) by contrasting the scaling behavior of the Helicity modulus $\Upsilon_x$ at the transitions. We  perform a finite size scaling analysis on $\Upsilon_x$ and the order parameter $M$, and compare their behaviors with those in KT transition. The results for two sets of parameters in the Quasi-ordered phase and the Ordered phase have been shown in Fig.~(\ref{fig:mmssize}).  

In the classical XY-model, the helicity modulus scales with the finite size $N$ of the system as
\be
\Upsilon_x(N) = \Upsilon_x(\infty)\Big(1 + \frac{1}{2} \frac{1}{ln N + C}\Big),
\ee
where  C is an undetermined constant~\cite{Weber1998}. At the KT transition point $K = K_c$, the helicity modulus has a jump $\Upsilon_x(\infty) K_c = 2/\pi$. Both the finite size scaling and the value at the jump have been verified \cite{Sudbo} at the Disordered to the Quasi-ordered transition. The behavior is quite different in the ordered phase. The stiffness $\Upsilon_x(N)$ in this transition already develops for $\alpha > \alpha_c$ at small sizes and remains unchanged with $N$. For $\alpha < \alpha_c$, $\Upsilon_x(N)$  decreases exponentially.

 The magnetization in the Quasi-Ordered KT phase is 0 in the limit $N \to \infty$. But the passage to this limit is very slow \cite{Bramwell1994}.
The finite size scaling is quite different at the Ordered state as shown in Fig.~(\ref{fig:mmssize}). While $M_{2D}$ decreases with $N$ at small $N$, it is consistent with  saturation at a finite value at large $N$, merging with the value of $M$. As discussed immediately after the definition of $M$ and $M_{2D}$ above, this is consistent with a truly Ordered state. 

\begin{figure*}[tbh]
\centering
\includegraphics[width=0.45\textwidth]{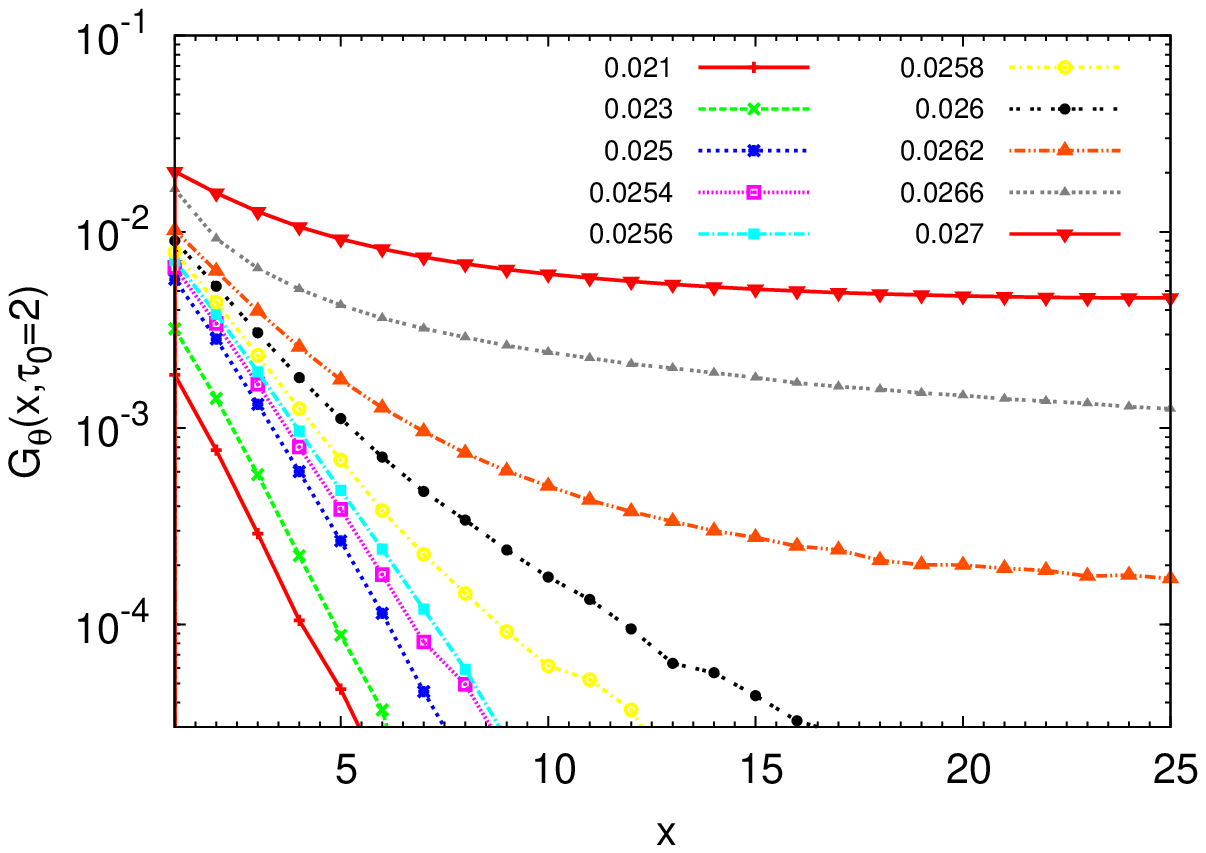}
\includegraphics[width=0.45\textwidth]{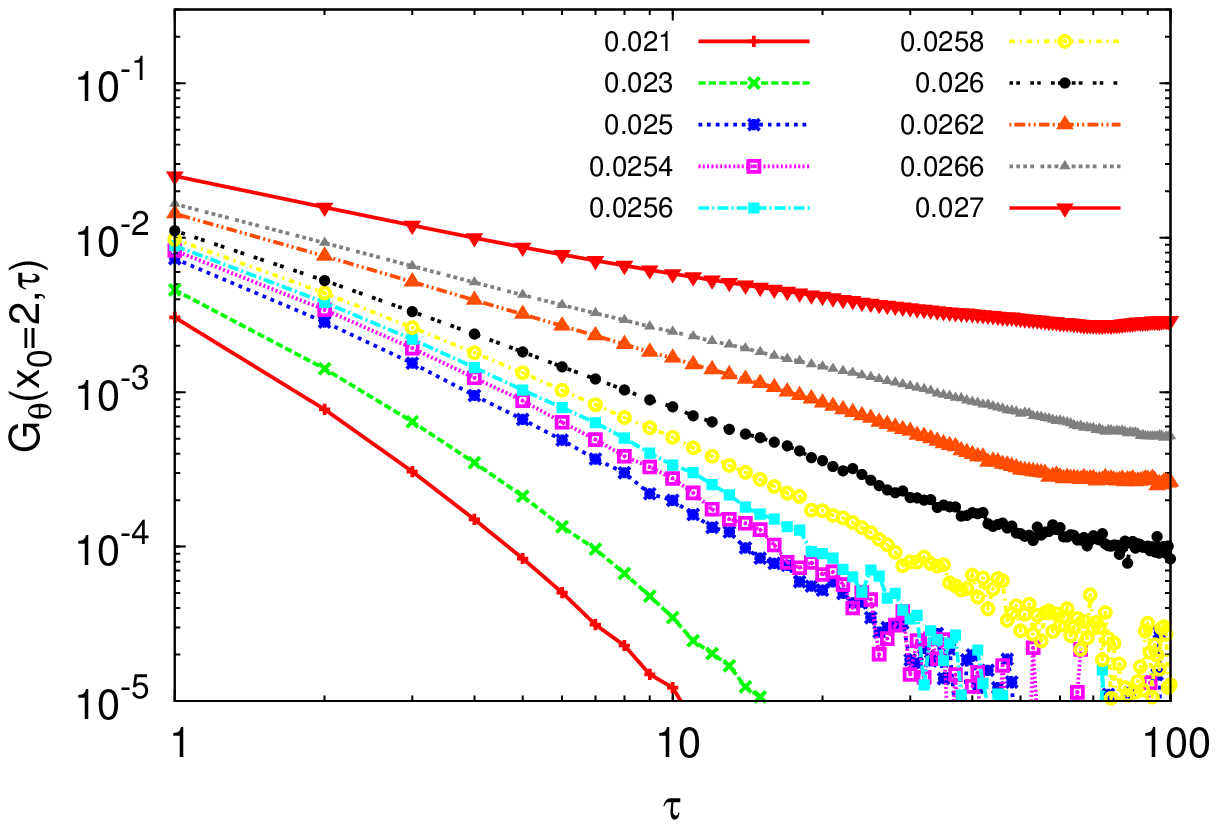}
\caption{The order parameter correlation functions $G_{\theta}(x,\tau)$ for transition from the Disordered phase to the Ordered phase. Parameters are the same as in Fig.~(\ref{fig:13static}). We show $G_{\theta}(x,\tau)$ as a function of $x$ for fixed $\tau=2$ (left panel) and as a function of $\tau$ for fixed $x=2$.  }
\label{fig:13cor}
\end{figure*}

\subsection{Scaling of the order parameter correlation functions}
\label{sec:cor_scale}

The most revealing results about the critical properties are of course obtained from the order parameter correlation functions.  It is seen in Fig.~(\ref{fig:13cor}) that there exists a separatrix in $G_{\theta}(x, \tau)$ for a fixed $x$ or for a fixed $\tau$ such that, for $\alpha < \alpha_c$ the asymptotic correlation $\to 0$ for large $\tau$, and for $\alpha > \alpha_c$, they tend to a constant value depending on $\alpha$. We present  scaling analysis of the order parameter correlation functions on the disordered side. 

\begin{figure*}[tbh]
\centering
\includegraphics[width=0.45\textwidth]{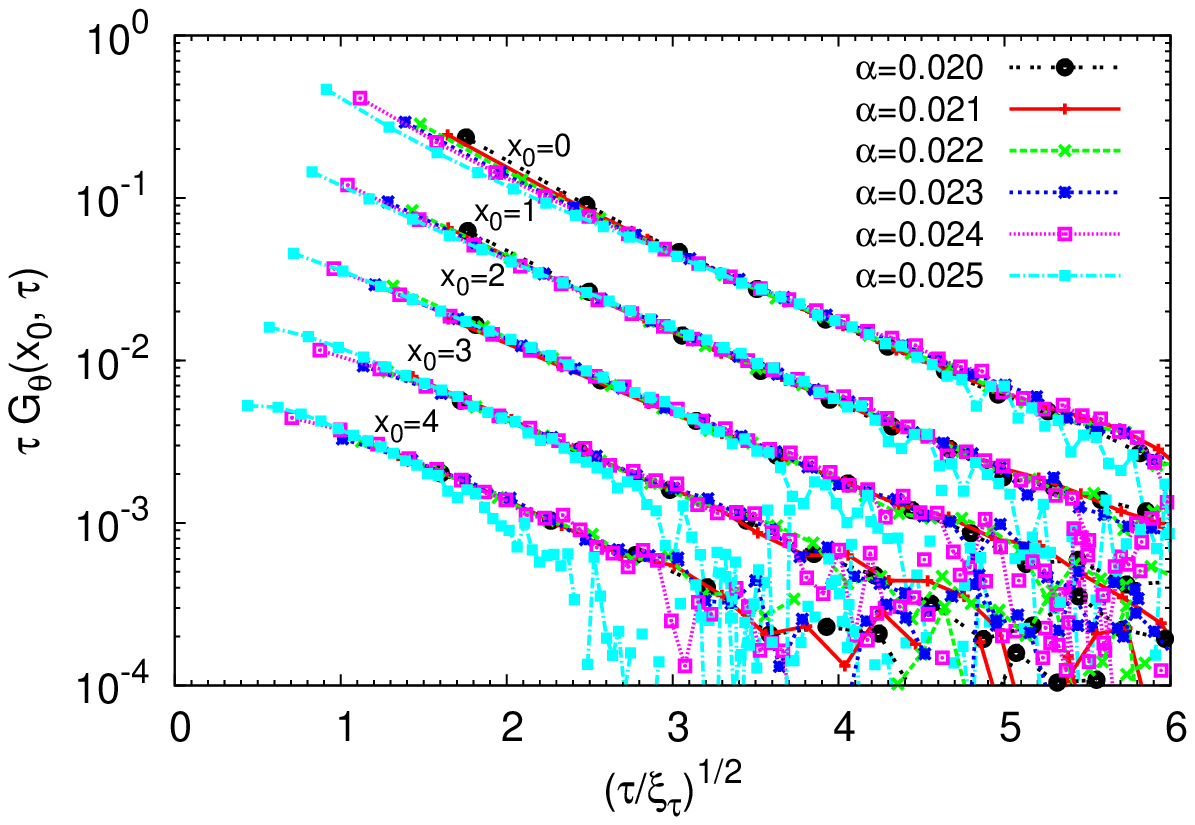}
\includegraphics[width=0.45\textwidth]{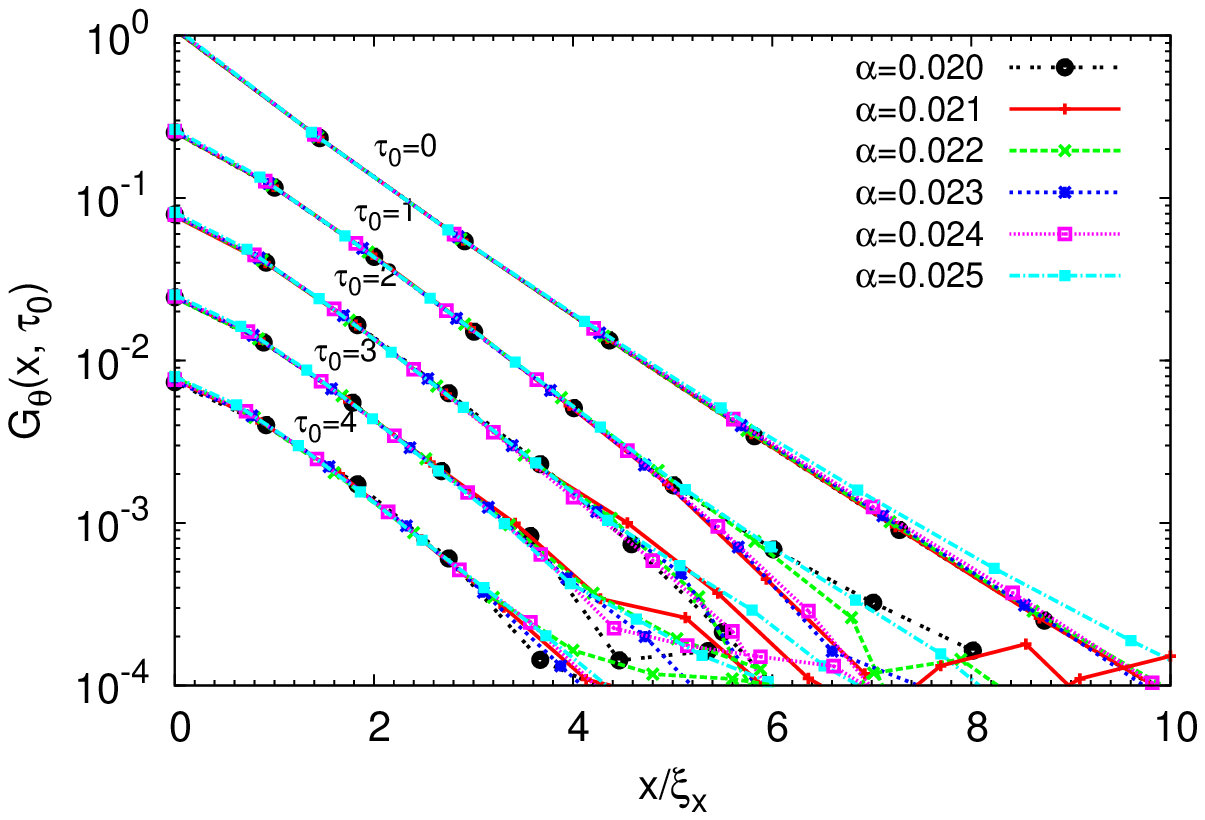}
\caption{Scaling analysis of the order parameter correlation function $G_{\theta}(x, \tau)$ for fixed $x=x_0$ (left panel) and for fixed $\tau=\tau_0$ (right panel) from the disordered side of the Disordered to Ordered phase transition as shown in Fig.~\ref{fig:13cor}.  In the left panel, we fit each curve of  $\tau G_{\theta}(x_0, \tau)$ with the form $ A_{\tau}(x) \exp[-(\tau/\xi_{\tau})^{1/2}]$, where the amplitude $A_\tau$ and the correlation length $\xi_\tau$ are fitting parameters adjusted for each $\alpha$ and $x$. In the right panel, we fit each curve of $G_\theta(x, \tau_0)$ with $A _x(\tau) \exp(-x/\xi_x)$ where $A_x(\tau)$ and $\xi_x$ are fitting parameters. The results of $\xi_{\tau}(x_0, \alpha)$ and $\xi_x(\tau_0, \alpha)$ are shown in Fig.~(\ref{fig:corrlength}). We find that $A_{\tau} \approx \tau_c \exp(-x/\xi_{0,x} )$ with $\tau_c \approx 0.12$ and $\xi_{0,x} \approx 1.0$,  and $A_x \approx (\tau_c/\tau) \exp[-(\tau/\xi_{\tau}(\alpha-\alpha_c))^{1/2}]$ with $\xi_\tau(\alpha-\alpha_c)$ given in Eq.~(\ref{eq:xitau}). It is expected that all curves of $\tau G_\theta(x_0, \tau)/A_{\tau}$ for difference $\alpha$ and $x_0$ collapse into a single curve $\exp(-t)$ with $t=(\tau/\xi_{\tau})^{1/2}$, which are plotted (for clarity, they are rescaled by a factor $10^{(x_0/2)}$ for different $x_0$). $G_\theta(x,\tau_0)/A_x$ as functions of $x/\xi_x$ are plotted in the same fashion. Because of the rapid decay of the correlation function in this range of $\alpha$, it has not been numerically possible to follow its behavior for larger $x$ and $\tau$. } 
\label{fig:13mcort}
\end{figure*}

We find that the leading asymptotic behaviors of $G_{\theta}(x, \tau)$ can be captured in the scaling form 
\be
\label{scaling}
G_\theta(x, \tau) = \frac{A}{\tau^{1+\eta_{\tau}}} e^{-({\tau/\xi_{\tau}})^{1/2}}\frac{1}{x^{{\eta}_x}}e^{-x/\xi_{x}}, 
\ee
where $\xi_{\tau}$($\xi_x$) are correlation lengths along temporal(spatial) directions, and $A$ is the amplitude. From the detailed results given in Appendix  \ref{sec:scalingform}, we determine that the anomalous exponent $\eta_{\tau} \approx 0$. We cannot determine the anomalous exponent $\eta_x$ reliably in the numerical calculations because even close to the critical point, where the temporal dependence fits the $1/\tau$ behavior, the spatial dependence continues to be exponentially decreasing as a function of $x$ up to more than 1/2 the largest sizes that we can numerically calculate, please see Fig. (\ref{fig:13cor}). Above that range, it appears to approach a constant, but could be consistent with a logarithmic ($\eta_x =0$) form. Some discussion of this issue is given in the concluding section. 

The correlation functions are shown for a few fixed $x$ as functions of $\tau$ in the left panel of Fig.~(\ref{fig:13mcort}) and for a few fixed $\tau$ as functions of $x$ in the right panel of the same figure. Fitting the correlation functions to the scaling form in Eq.~(\ref{scaling}), we determine $\xi_{x}$ and $\xi_\tau$ for each $\alpha$.  We show them as functions of $\alpha-\alpha_c$ in Fig.~(\ref{fig:corrlength}). More details are provided in Appendix \ref{sec:scalingform}.
In the fluctuation regime not too close to the critical point  in the disordered side, for $(\alpha_c - \alpha)/\alpha_c \gtrsim 0.1$ with $\alpha_c \approx 0.0260$, we observe that in the parameter range shown, $\xi_{\tau}$ increases by a decade when $\alpha \to \alpha_c$ while $\xi_x$ remains relatively unchanged $\xi_x \approx \xi_{0,x} \approx 1.0$, i.e, a lattice constant. In this range of $\alpha$, the behavior of $\xi_\tau$ is consistent with 
\be
\xi_{\tau}(\alpha - \alpha_c) = \tau_c e^{ a\sqrt{ \alpha_c /(\alpha_c - \alpha)}},
\label{eq:xitau} 
\ee
where $a$ is a constant of O(1). This relation, as well as the leading behavior of the correlation function $G_\theta(x, \tau)$
\be
\label{leadscaling}
G_{\theta}(x, \tau) \approx \frac{\tau_c}{\tau} e^{-({\tau/\xi_{\tau}})^{1/2}}e^{-x/\xi_{0,x}},
\ee
are identical to those derived analytically \cite{Aji2007} (the dependence on $\xi_\tau$, $e^{-({\tau/\xi_{\tau}})^{1/2}}$, has not been derived explicitly).  $\tau_c$ is the short-time cutoff scale. It was also derived that, within factors of O(1), $\tau_c =(1/\sqrt{K_0E_c})/ \Delta \tau=1/\sqrt{K/K_{\tau}}$. For the parameters chosen,  $1/\sqrt{K/K_{\tau}}$ = 0.16, while the numerically obtained value is $\tau_c \approx 0.12$.

\begin{figure*}[tbh]
\centering
\includegraphics[width=0.48\textwidth]{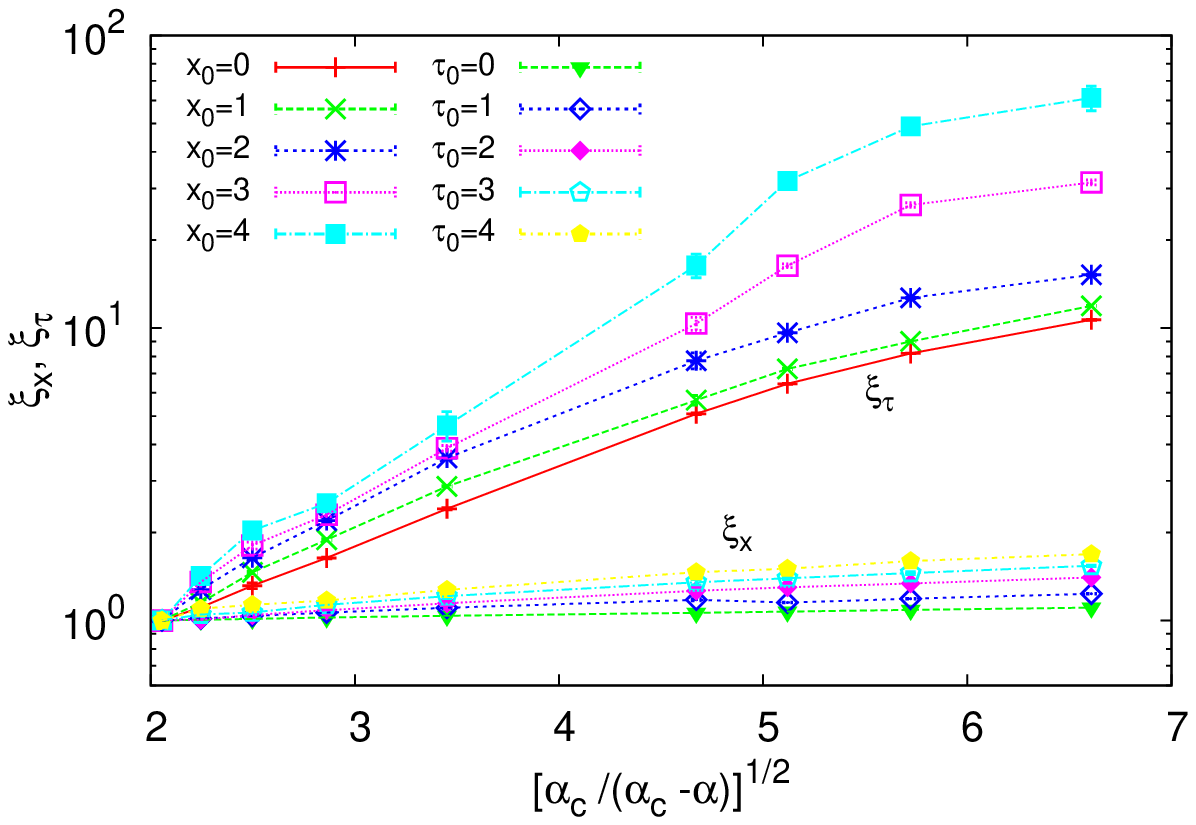}
\includegraphics[width=0.48\textwidth]{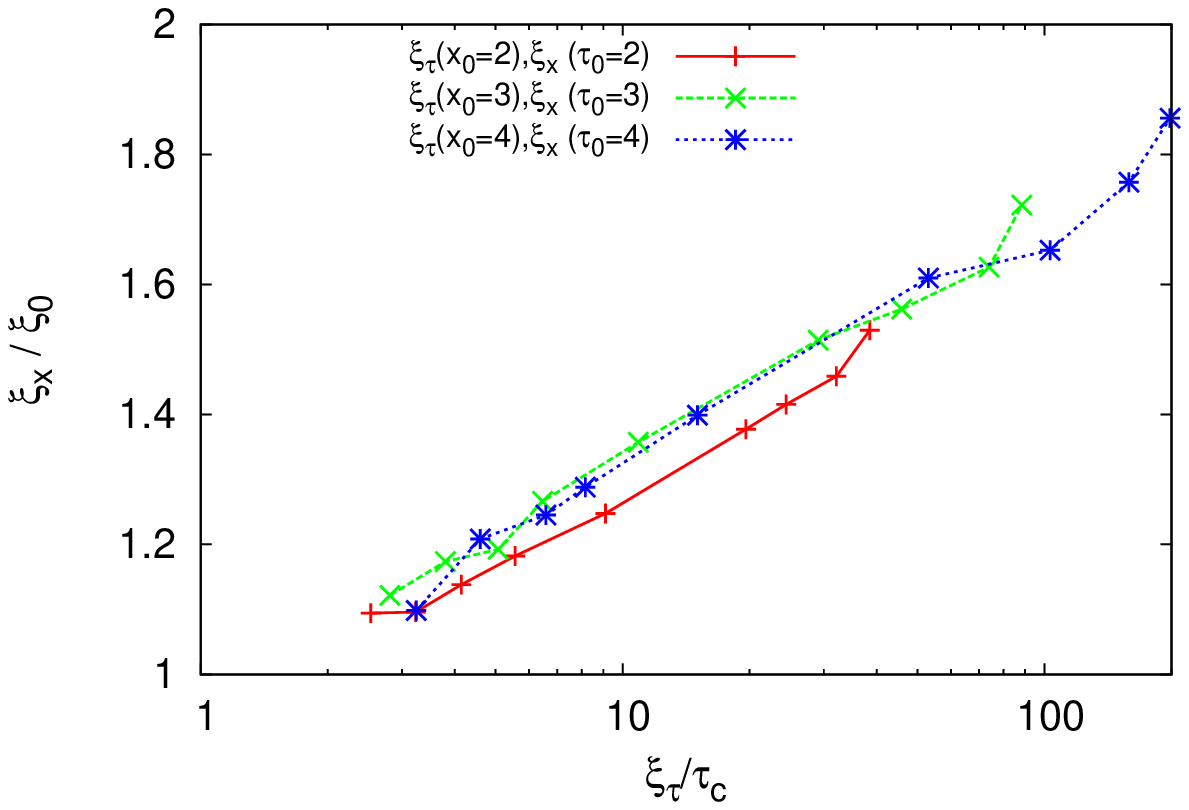}
\caption{ The left panel shows $\xi_x$ and $\xi_{\tau}$ as functions of $[\alpha_c /(\alpha_c - \alpha)]^{1/2}$. They have been rescaled to their respective values at $\alpha=0.020$ ($[\alpha_c /(\alpha_c - \alpha)]^{1/2}\approx 2$). The error bar is typically smaller than the symbol size; it increases for large $\xi_x$, i.e. closer to the critical point - note the logarithmic scale. For $x_0=0$,  $\xi_{\tau}$ can be fitted as $\tau_c \exp [0.62\sqrt{\alpha_c /(\alpha_c - \alpha)}]$; the numerical coefficient in the exponent changes to about 1 for $x_0 = 4$. The right panel shows the relation between $\xi_x(\alpha)$ and $\xi_{\tau}(\alpha)$.  We find that $\xi_x/ \xi_0 \sim \ln (\xi_\tau/\tau_c)$. This relation appears to become independent of $x$ and $\tau$ at large  $x$ and $\tau$. Finite size effects do not permit a detailed exploration beyond $\xi_\tau/\tau_c \approx 70$. }
\label{fig:corrlength}
\end{figure*}

However, for $(\alpha -\alpha_c)/\alpha_c \lesssim 0.1$ on the disordered side, there are  deviations from Eqs.~(\ref{eq:xitau}) and (\ref{leadscaling}) . For example, we notice in Fig.~(\ref{fig:13cor}), a crossover from an exponential to a power law behavior in the spatial correlation as $\alpha \to \alpha_c$ before going to a constant value on the ordered side, consistent with true long-range order. As shown in the left panel of Fig. (\ref{fig:corrlength}), $\xi_{x}$ also increases when $\alpha \to \alpha_c$, though at a much slower rate compared to $\xi_{\tau}$. Their monotonic growth suggests scaling one with respect to the other. In the right panel of the same figure,  we show that within our numerical capabilities that
\be
\xi_x/\xi_{0,x} \approx  \ln{(\xi_{\tau}/\tau_c)},
\ee
i.e, the spatial correlation length is consistent with growing as the logarithm of the temporal correlation length \cite{footnote3}.  This means that the dynamical critical exponent is $z = \infty$. One should expect, as is consistent with Fig.~(\ref{fig:corrlength}), transients for $x \lesssim \xi_x$ and $\tau \lesssim \xi_{\tau}$ approaching the forms given above. In Appendix B, we show that within the numerical precision of our results, the relation $\xi_x \propto \xi_{\tau}^{1/8}$, rather than the logarithmic relation is allowed. Exponents larger than 1/8 or $z< 8$ are disfavored.

These properties, as well as what has been calculated above about the vortices, appear to be consistent with the suggestion \cite{Aji2007} that when warps begin to freeze, spin-waves might develop a gap so that the vortices also order (however, this has not been explicitly derived). The approximate correlation function (\ref{leadscaling}) is a separable function of space and time, and so is the final form of the correlation function (\ref{scaling}). However, a weak $\tau$ dependence $\propto \ln{(\tau/\tau_c)}$ cannot be excluded in $\xi_x$ very close to criticality. This question can only be settled by further analytical calculations, possibly by a proper renormalization group calculation of the effect of the last term in Eq.~(\ref{topomodel}). 

From the results here as well as from Ref.~[\onlinecite{Aji2007}], the phenemenological expression \cite{Varma1989} for quantum-critical fluctuations acquires a cross-over towards purely quantum-fluctuations below a cross-over temperature $T_x \approx \xi^{-1}_{\tau}$. This presents an essential singularity at the critical point in terms of the tuning parameter of the transition, $\alpha-\alpha_c(K, K_{\tau})$.

We have presented results for the correlation lengths as a function of $(\alpha-\alpha_c)$. As is evident from the phase diagram, $\alpha_c$ depends on $K$, and (not explored in this paper) on $K_{\tau}$, as well. Away from the meeting point of the three transitions, $\alpha_c$ depends smoothly on $K$. Therefore, we should expect that for fixed $\alpha$, the change of correlation length is the same function of $(K-K_c)$ as it is of $(\alpha-\alpha_c)$ for a fixed 
$K$. However, this point could benefit from further study.

We provide here the form of the correlation functions in frequency-momentum space (assuming $\eta_{\tau} = \eta_x =0$ ) which is convenient to compare with experiments as well as to calculate scattering of fermions from such fluctuation. The Fourier transform from the imaginary time-dependence to real frequency is described in Appendix \ref{sec:Fouriertransform}, where we show that the final result can only be obtained numerically. We find that the numerical results can be approximately fitted by the form
\be
\label{gthetaomegak1}
\text{Im} G_{\theta}(q, \omega) &\approx& G_0 \frac{1}{q^2 + \kappa_x^2} ~\rho(\omega, T, \kappa_{\tau}),
\ee 
\be
\label{gthetaomegak2}
\rho(\omega, T, \kappa_{\tau}) &\to &  \frac{\omega}{2\sqrt{T^2 + 0.4 \kappa_{\tau}^2}}, \text{for} ~\beta{\omega} \to 0,  \\ \nonumber
& \to &\frac{1}{4}(1+3 e^{- [\beta \kappa_{\tau}/2]^{1/2}}), ~ \text{for} ~ \beta\omega \gg 1. 
\ee
Here $\kappa_x = \xi^{-1}_x(\alpha, K, K_{\tau}), \kappa_{\tau} = \xi^{-1}_{\tau}(\alpha, K, K_{\tau})$ and $G_0$ measures the integrated strength of the fluctuations. The following features of  $\text{Im} G_{\theta}(q, \omega)$ are especially noteworthy. (i) It is a separable function of $q$ and $\omega$. (ii) In the critical region, i.e. $T/\kappa_{\tau} \gg 1$,   $\rho \propto \tanh(\frac{\omega}{2T})$. It should be noted that $\kappa_{\tau}$ is such a slow function of $(\alpha-\alpha_c)/\alpha_c$ [see Eq.~(\ref{eq:xitau})], that the quantum critical region may be visible over a very wide region of parameters on the disordered side. (iii) The low frequency part is cut-off for $T/\kappa_{\tau} \ll1$ with $\kappa_{\tau}$ replacing $T$. (iv) For large $\beta\omega$, there is a rapid decrease of the correlation function with $\kappa_{\tau}/T$. Ultimately, there is a ultra-violet cut-off of the frequency $\omega_c = \tau_c^{-1}$. In any given experimental systems, there may be cut-offs not included in the XY model, for example the Fermi-energy if it is similar to or smaller than $\omega_c$. 

\section{The effect of four-fold anisotropy}
\label{sec:h4}  

\begin{figure*}[tbh]
\centering
\includegraphics[width=0.8\textwidth]{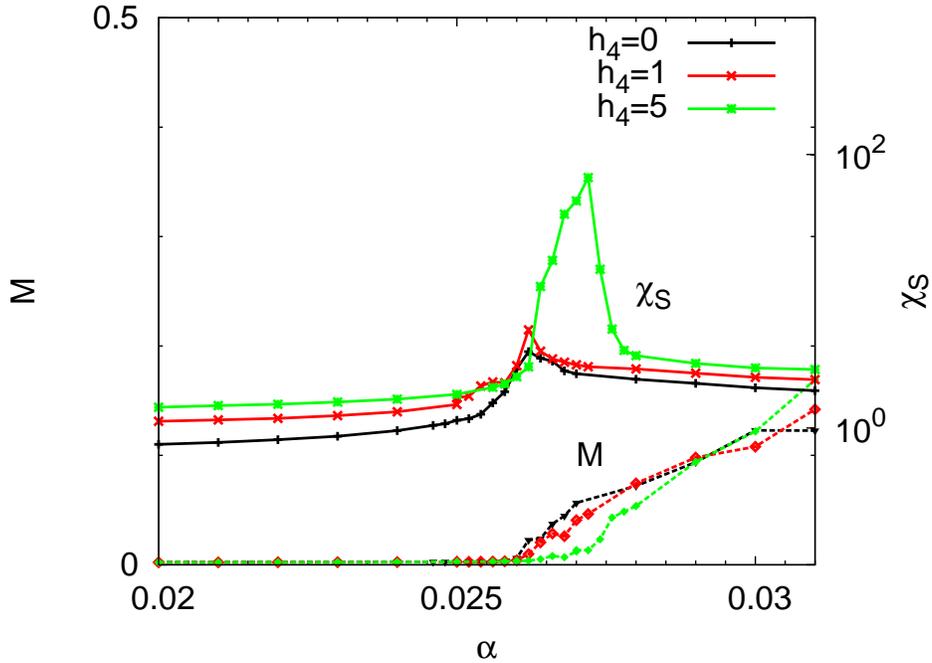}
\caption{ $\chi_{S}$ and $M$ for different $h_4$. Here,  $K=0.4$, $K_{\tau}=0.01$ and $\alpha$ is varied. The system size is  kept the same, $N=50$ and $N_{\tau}=200$. }
\label{fig:h4static}
\end{figure*}

\begin{figure*}[tbh]
\centering
\includegraphics[width=0.45\textwidth]{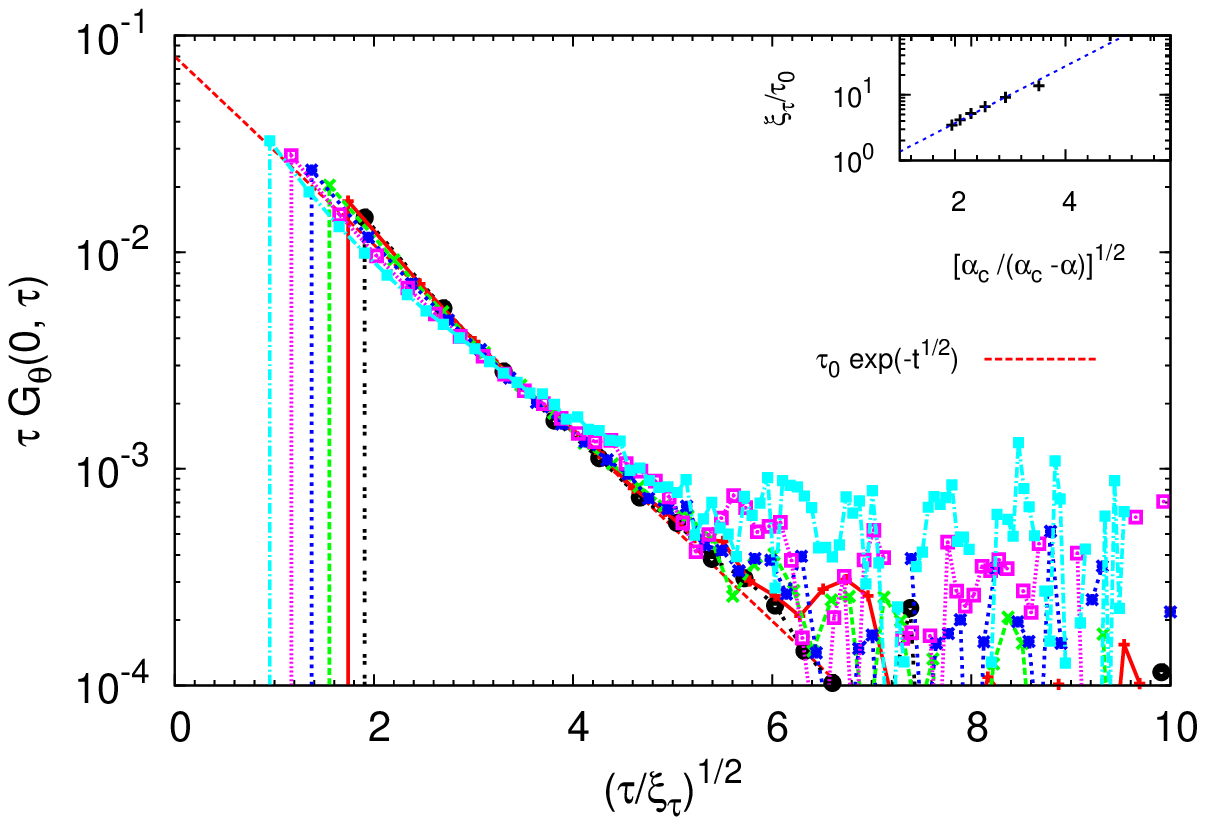}
\includegraphics[width=0.45\textwidth]{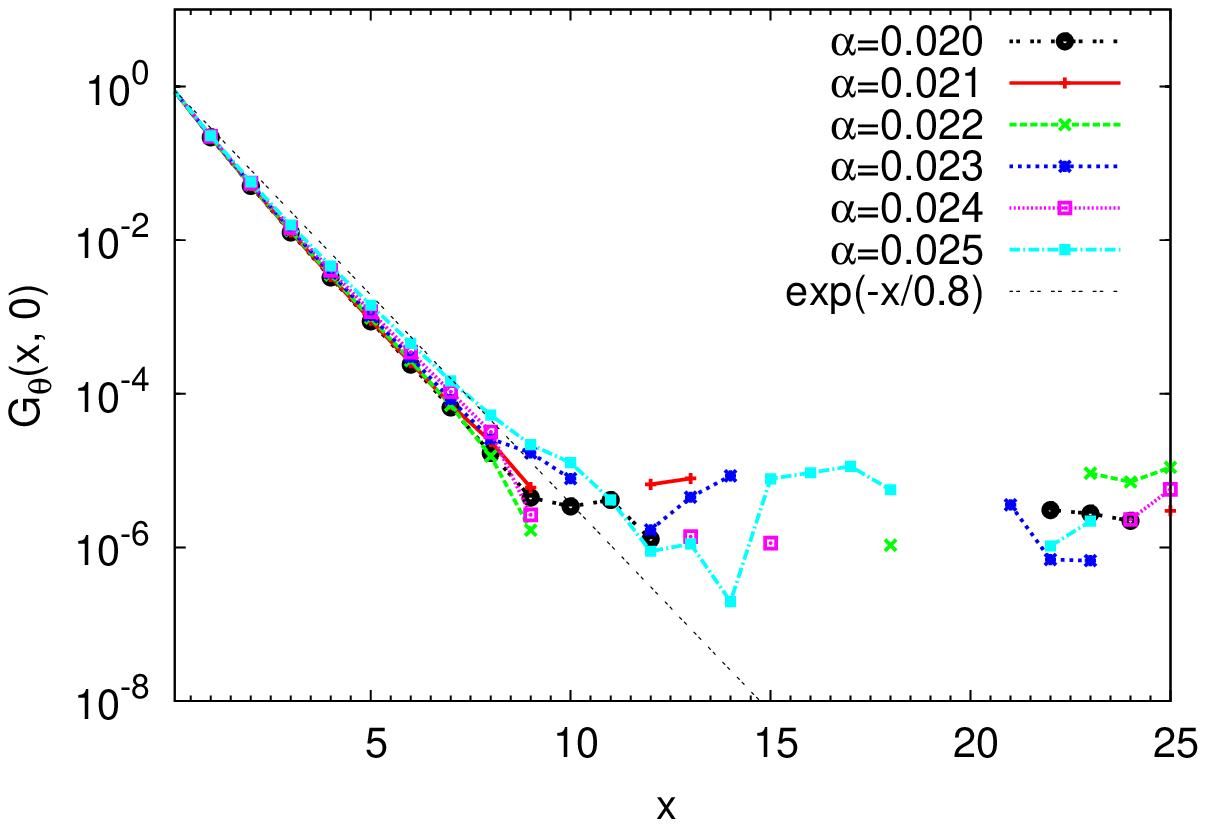}
\caption{Order parameter correlation functions for $h_4 =5$. The left panel show scaling analysis of the ${\bf x}=0$ spin correlation function $G_{\theta}(0, \tau)$ from the disordered side of the transition. The fitting curve is similar to that in $h_4=0$ case,  $G_{\theta}(0, \tau)=(\tau_0/\tau)\exp[-\sqrt{\tau/\xi_{\tau}}]$ with $\tau_0\approx 0.08$.  The inset shows $\xi_{\tau}/\tau_0$ as a function of $\sqrt{\alpha_c /(\alpha_c - \alpha)}$. One finds that $\xi_{\tau}(\alpha)/\tau_0 \approx 0.5\exp [\sqrt{\alpha_c /(\alpha_c - \alpha)}]$, with $\alpha_c =0.0272$.  The right panel shows equal-time spin correlation functions $G_{\theta}(x, 0)$ as functions of $x$.  For $x \lesssim 10$, they have also the same form as in $h_4=0$ case $G_{\theta}(x, 0)=\exp(-x/\xi_x)$ with $\xi_x \approx 0.8$ .}
\label{fig:h4mcor}
\end{figure*}

We now turn on the four-fold anisotropic field $h_4$ in the Monte-Carlo simulation to study its effect. In the classical XY model, 4-fold anisotropy is marginally irrelevant \cite{Jose}. In the quantum model, it has been argued \cite{Aji2007} to be irrelevant. When $h_4\to \infty$, XY spins become two Ising variables, as in Ashkin-Teller model. We focus on the transition from the Disordered phase to the Ordered phase, by choosing $K=0.4$, $K_{\tau}=0.01$ and tuning $\alpha$ for transitions for different $h_4$. We find that the transition persists and all quantities  have similar properties across the transition as in $h_4=0$ case. In Fig.~(\ref{fig:h4static}), we compare $\chi_S$ and $M$ for three different values of $h_4=0, 1, 5$. We find that up to $h_4=1$, the properties are almost the same as in $h_4=0$. In $h_4=5$, we notice that $\alpha_c$ has been shifted to $0.0272$, and the peak in $\chi_S$ is sharper. $M$ increases more rapidly.  

We further show the scaling results of the spin correlation functions for $h_4=5$ in Fig. ~(\ref{fig:h4mcor}). We find similar behaviors as in $h_4=0$ case, which indicates that the transition is also of the local-critical type. 

\section{Discussion}
\label{sec:conclusion}

In this paper, the properties of the dissipative quantum XY model have been investigated by Monte-Carlo simulations to verify and extend the analytical calculations in Ref.~[\onlinecite{Aji2007}] and the previous Monte-carlo simulations in Ref.~[\onlinecite{Sudbo}]. We have found properties consistent with local quantum-criticality of the form proposed in Ref.~[\onlinecite{Varma1989}] and derived in Ref.~[\onlinecite{Aji2007}] with a crossover in time/temperature to the disordered quantum state with some important modifications. Very importantly, we have also found a new result: a spatial correlation length which however varies very slowly, consistent with logarithmically, with the temporal correlation length. It is hoped that this result can also be derived analytically, as also the anomalous exponent $\eta_x$. 
  
We re-emphasize that the conclusions based on numerical results at finite $N$ and $N_{\tau}$ can at most be highly suggestive. In Appendix \ref{xixxit-powerlaw} we show that a dynamical critical exponent of 8 fits the data as well as $\infty$.  The separability of the spatial and temporal dependence of the correlations, which is a novel feature of the results, depends on the numerical capabilities in which the results are obtained. It is possible that very close to the critical point, $(\alpha_c - \alpha)/\alpha_c \lesssim 10^{-2}$, the results could be different. This region is affected by the finite-size, or finite temperature effect, where classical dynamics dominates. Critical slowing-down could also be a contributing factor. Such issues are best addressed by analytic methods, to which the present results serve as a guide. However, we can be fairly certain that over the range which is quite close to a critical point, the spatial correlations vary very slowly compared to the temporal correlations and the two are separable, and that the disordered to ordered transition is driven by freezing of the warps with the vortices freezing when the warp correlations become sufficiently long. These results are in the range in which experiments are usually done.
  
We should also stress that most of the study on the correlation functions is on the disordered side of the quantum-critical point. Some comments may be worthwhile on the ordered side. The ordered side for the problem studied has the properties of the model without dissipation, i.e. it is the ordered phase of the 3d-XY type. Some of our preliminary results indicate that as $K_{\tau}$ is increased, the region of the Quasi-ordered phase decreases in the $K-\alpha$ plane. This is in agreement with the fact that  when $\alpha \to 0$, the transition as tuned by the ratio of $K/K_\tau$ is of the 3D XY type, in which the correlations are expected to be a function of the co-ordinate $(x^2 + v^2 \tau^2)^{1/2}$, where $v^2$ is given dimensionally in the third term of Eq.~(\ref{G}) by $K /K_\tau$. The results in this paper on the disordered side suggest that $v^2$ scales near the transition in an interesting way. In the critical region on the disordered side, it vanishes when away from the critical point, indicating that the long-range correlations develop only in time. On the ordered side, it acquires a finite value.  We also know that the theory is non-analytic as $\alpha \to 0$. Properties in the $K_\tau - \alpha$ plane are subjects of further study. 
 
It is not the purpose of this paper to discuss the experiments which may be related to the findings here. But a few comments about future directions in relation to both theory and experiments may be worth-while.

The dissipative quantum XY model was first proposed \cite{chakra86, Fisher86} in connection with the superconductor to insulator transition in thin superconducting films \cite{expts}. Quite correctly, the transition as a function of dissipation was proven. But the fluctuation spectra in various calculations \cite{prevtheor} in two dimensions were not obtained in a controlled manner and do not agree with the results presented here and in Ref.~[\onlinecite{Aji2007}]. (However, the results for the one-dimensional array of Josephson junctions in a dissipative environment \cite{rafael} are closely related to the results here and in Ref.~[\onlinecite{Aji2007}].) Nor do the results of these calculations give the rich phase diagram found in [\onlinecite{Sudbo}] and here, which is suggested by re-expression of the model in terms of warps and vortices. It would be interesting to think of how experiments might discover the different phases in a superconducting thin film.  We are also not aware of experiments to probe the fluctuation spectra at the superconductor to insulator transitions. This would also be very interesting to pursue, possibly  by studying fluctuations across a Josephson junction to a three-dimensional superconductor below its transition temperature. To fully understand such possible experiments, the present work should be extended to include (the equivalent of) a magnetic field. 

The dissipative quantum XY model (with four-fold anisotropy) has also been proposed \cite{cmv} as a model for the observed order \cite{bourges} in the under-doped region of the cuprates. The phenomenological quantum-critical fluctuations, which have been successful in explaining the diverse anomalies in the strange metal region of these compounds, have now been proven to be the property of the fluctuations of the observed order. It is remarkable that some of the same anomalies observed in the cuprates in this region also occur in the AFM quantum-critical region of some of the heavy-fermions and in the Fe-based superconductors. This has led to the inquiry and the conclusion \cite{cmv-afm} that the criticality of a simple model of itinerant AFM is also described by the dissipative XY model.

\begin{acknowledgements}                                                                              
We thank Asle Sudb{\o} for several conversations and correspondence about the work in Ref.~[\onlinecite{Sudbo}], which were very important in the present work. We wish to acknowledge very helpful discussions with Vivek Aji.  Some calculations by Wei Yan and Jie Lou at a preliminary stage of this work are also acknowledged. This work was partly supported by the National Science Foundation under grant NSF-DMR 1206298.
\end{acknowledgements}                                                                                
                               
\appendix
\section{The scaling form of the order parameter correlation functions}
\label{sec:scalingform}

\begin{figure*}[tbh]
\centering
\includegraphics[width=0.45\textwidth]{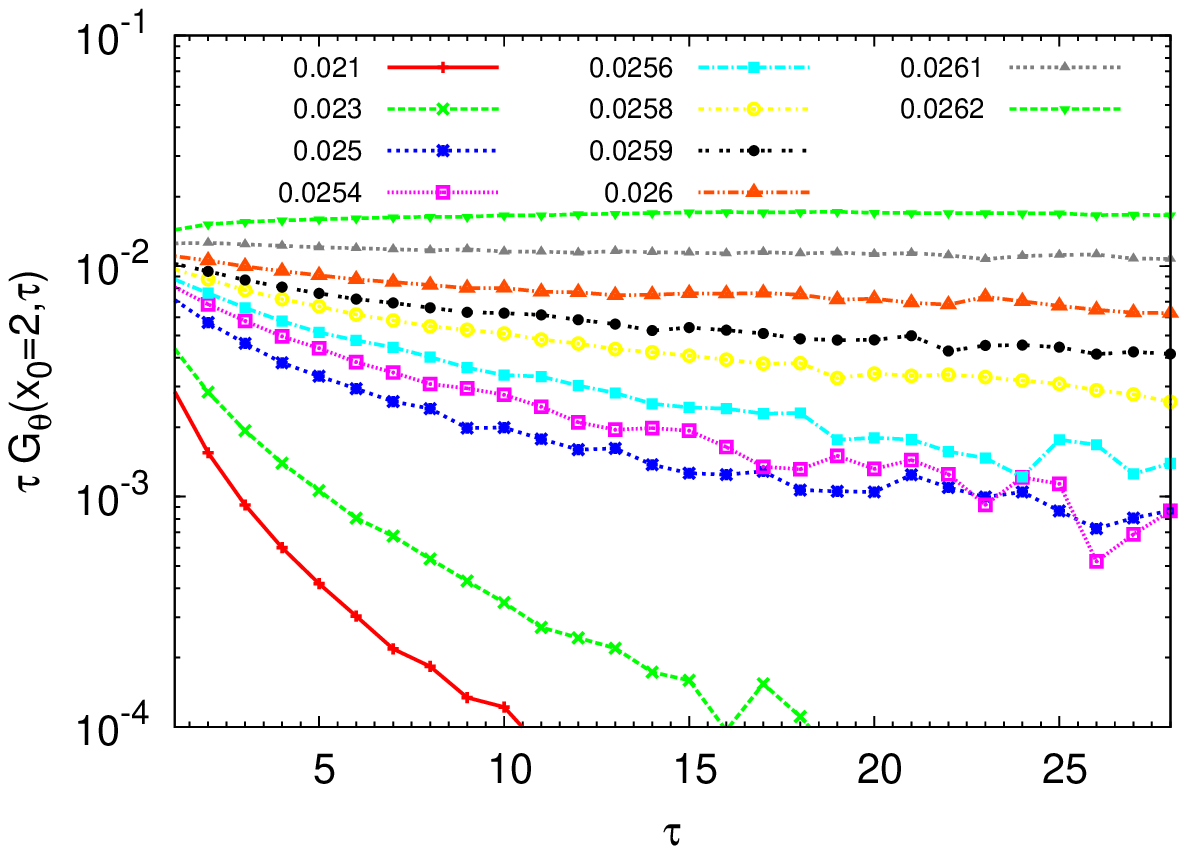}
\includegraphics[width=0.45\textwidth]{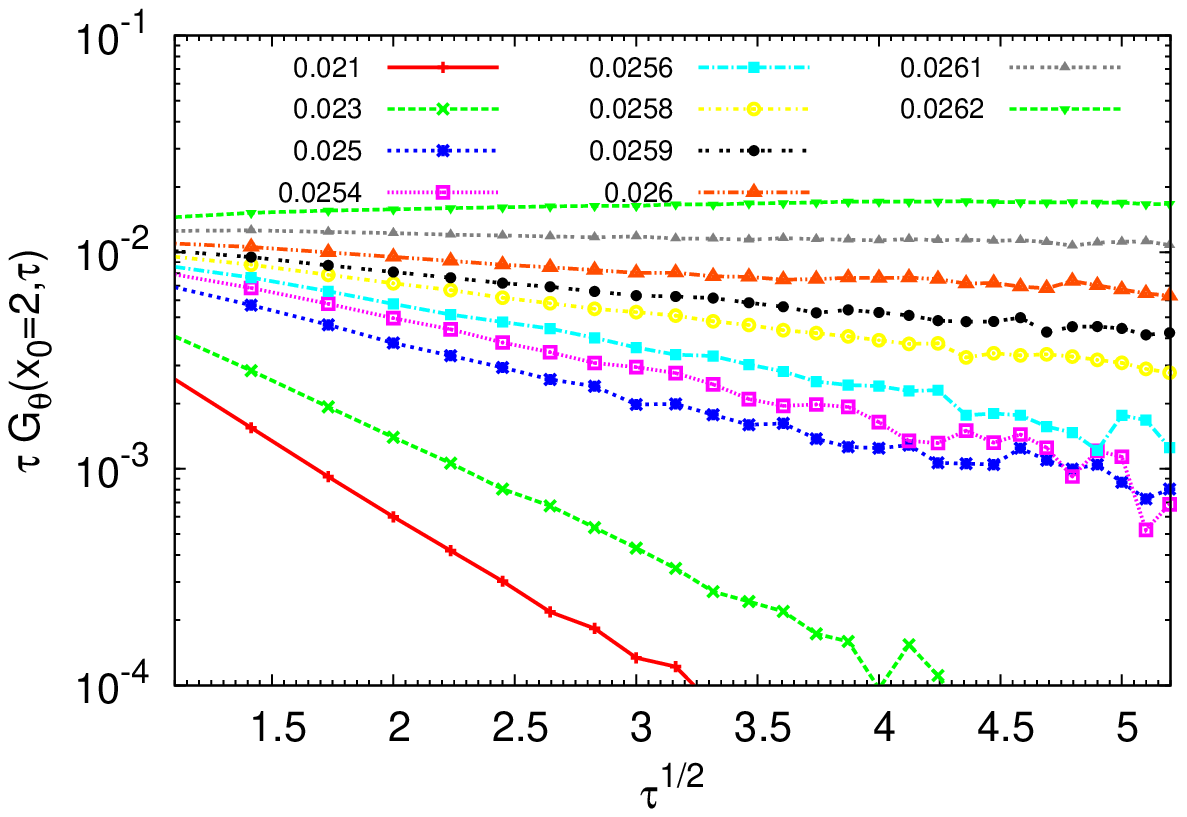}
\caption{ $\log[\tau G_{\theta}(x_0=2,\tau)]$ as functions of $\tau$ (left panel) for selected $\alpha$s in transition from the Disordered phase to the Ordered phase. Other parameters are the same in Fig. ~(\ref{fig:13static}).  The right panel shows them as functions of $\tau^{1/2}$ instead. }
\label{spincortimest}
\end{figure*}

We present here results with a much finer variation in $\alpha$ so as to place better bounds on our results. We first discuss the correlations along temporal direction with $x$ fixed at a small value, which take the general scaling form 
\begin{equation}
G_{\theta} (x_0, \tau) \sim \frac {1}{\tau^{1+\eta_{\tau}}} e^{- (\tau/\xi_\tau)^p},
\end{equation}
where $\xi_\tau \to \infty$ approaching the critical point. In practice, fitting $\eta$ and $p$ simultaneously lead to uncertainties. The analytical study shows the anomalous scaling dimension for temporal correlation is $\eta_\tau=0$. Fig.\ref{spincortimest} shows how the results fit into this form, by plotting $\tau G_{\theta}(x_0=2,\tau)$ as functions of $\tau$. Indeed, we find that near $\alpha_c$,  $\tau G_{\theta}(x_0=2,\tau)$ become almost a constant.  Another systematic check is that, very close to $\alpha_c$ and $\tau \ll \xi_{\tau}$, $G_{\theta} (x_0, \tau) \sim 1/\tau^{\eta}$, with only parameter to fit.  Fittings to $\alpha=0.026, 0.0261, 0.0262$ yields $\eta = 1.12\pm 0.03, 1.03\pm0.04, 0.94\pm 0.03$, respectively. We therefore determine $\eta_{\tau} \approx 0$.  Using this result, we can decide  $p$. A comparison between fit to the Monte-Carlo data with $p=1$ and $p=1/2$ is shown in Fig. \ref{spincortimest}. This shows that   $p=1/2$ is much preferred over $p=1$.   

From Fig.~\ref{fig:13mcort}, it is easy to see the exponential fall off of the spatial correlations characterized by the spatial correlation length $\xi_x$ and determine its dependence on $(\alpha-\alpha_c)$. But it has proven harder to determine the scaling dimension $\eta_x$ of the spatial dependence at criticality, as already discussed in the paper.

\section{The relation between $\xi_x$ and $\xi_\tau$: power-law fitting}
\label{xixxit-powerlaw}

\begin{figure*}[tbh]
\centering
\includegraphics[width=0.48\textwidth]{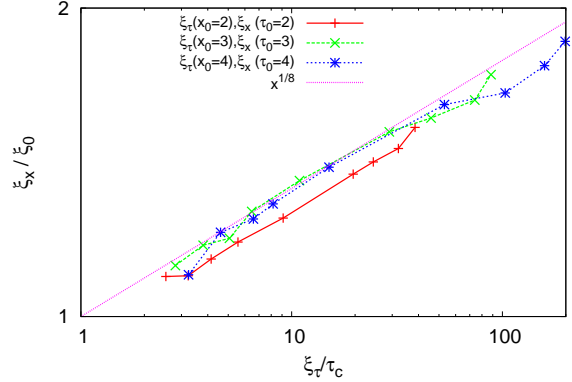}
\caption{ $\xi_x(\alpha; \tau_0)/\xi_0$ as functions of $\xi_{\tau}(\alpha; x_0)/\tau_c$ in log-log scale.  }
\label{fig:corrlengthlog}
\end{figure*}

In Fig.~\ref{fig:corrlengthlog}, we show the relation between $\xi_x$ and $\xi_\tau$ could also be fitted as $\xi_x \sim \xi_{\tau}^{1/8}$, or a dynamic exponent $z\approx 8$. An equally good fit has been shown to a logarithmic form (see Fig~\ref{fig:corrlength}). On aesthetic grounds, we may choose the latter.

\section{Spectral function of the order parameter correlation function}
\label{sec:Fouriertransform}

The correlation function is in a separable form of $\tau$- and $x$- dependent terms. The Fourier transform from spatial space to momentum space (${\bf q}$) is straightforward: for 2D, $\exp(-x/\xi_x)$ is transformed to $1/(q^2 + \xi_x^{-2})$.  Here we provide the details on Fourier transform from the imaginary time variable $\tau$ to the real frequency variable $\omega$ of the function in Eq.~(\ref{scaling}), with $\eta_{\tau} =0$. A bosonic correlation function in imaginary time $G_b(\tau)$ is related to its spectral function $\rho_b (\omega) = -(1/\pi) \text{Im} G_b(\omega+i0^+)$ by 
\begin{equation}
G_b(\tau) = \int_{-\infty}^{\infty} \frac {e^{-\tau \omega}} {1-e^{-\beta \omega} } \rho_b(\omega) d \omega,
\end{equation}
 for $0\le \tau \le \beta$. Setting $t  = i (\tau -\beta/2)$, we have
 \begin{equation}
 G_b(-it +\beta/2) = \int_{-\infty}^{\infty} \frac { e^{i t \omega}} {2\sinh (\beta \omega/2) } \rho_b(\omega) d \omega
 \end{equation}
or 
\begin{equation}
 \rho_b(\omega)  = \frac { \sinh (\beta \omega/2) }{\pi} \int_{-\infty}^{\infty}  G_b(-it +\beta/2)  e^{-i t \omega} d\,t.
 \end{equation}
 
 \begin{figure}[tbh]
\centering
\includegraphics[width=0.8 \columnwidth]{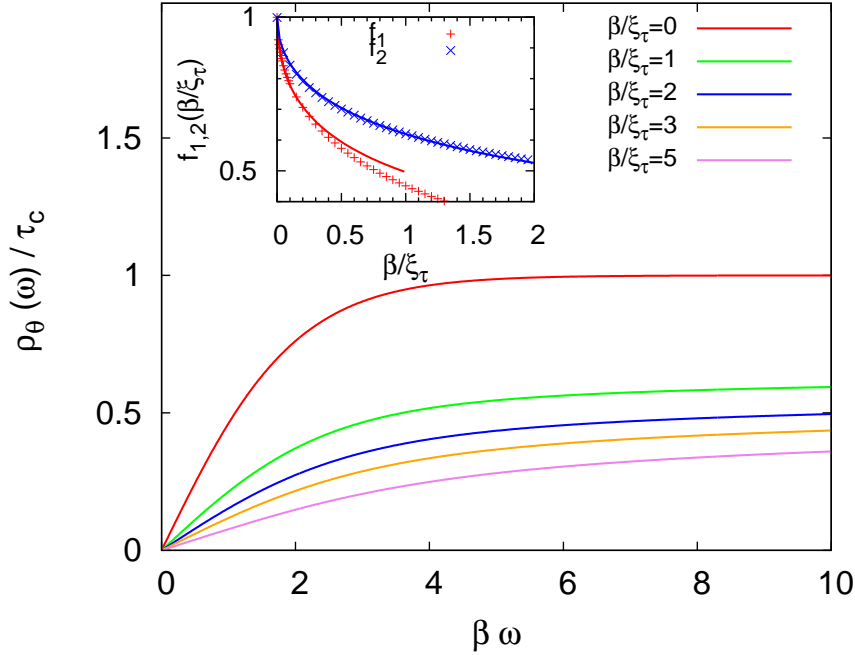}
\caption{ $\rho_\theta(\omega)/\tau_c$ as functions of $\beta \omega$  for selected values of $\beta/\xi_\tau$. When $\beta \omega \ll 1$,  $\rho_\theta(\omega)/\tau_c \to  f_1(\beta/\xi_\tau) \beta \omega/2$. When $\beta \omega \gg 1$, $\rho_\theta(\omega)/\tau_c\approx f_2(\beta/\xi_\tau)$.  The functions $f_1$ and $f_2$ are shown in the inset.  Their functional forms in the range of $\beta/\xi_\tau$ shown, can be fitted as $f_1(\beta/\xi_\tau) =1/[1+0.42 (\beta/\xi_\tau)^{1/2}]^2$, and $f_2(\beta/\xi_\tau) = \frac{1}{4}(1+3 e^{-[\beta/(2\xi_\tau)]^{1/2}})$, which are shown in solid lines. }
\label{fig:spectral}
\end{figure}

We rewrite the time-dependent part of the order parameter correlation into a periodic form
\begin{equation}
G_\theta(\tau) = \frac {\pi \tau_c}{\beta \sin (\pi \tau /\beta)} \left( e^{- (\tau/\xi_\tau)^{1/2}} + e^{- [(\beta-\tau)/\xi_\tau]^{1/2}} \right), 
\end{equation}
which is also particle-hole symmetric $G_\theta(\beta-\tau) =G_\theta(\tau) $, and therefore
\begin{eqnarray}
 \rho_\theta(\omega)  &=& 2 \tau_c \sinh \frac{\beta \omega}{2} \int_{0}^{\infty}  \frac {\cos\beta \omega t}{ \cosh (\pi t) } e^{- (\beta/\xi_\tau)^{1/2} [(1/2)^2+t^2]^{1/4} \cos[(\frac 12 \tan^{-1} 2t)]} \nonumber \\
 && \times 
 \cos\{(\beta/\xi_\tau)^{1/2} [(1/2)^2+t^2]^{1/4}\sin (\frac 12 \tan^{-1} 2t)\} d\,t.
 \end{eqnarray} 
This integral can only be evaluate numerically. Results as a function of $\beta\omega$ for several $\beta/\xi_\tau$  are shown in Fig.~\ref{fig:spectral}. For, $\beta/\xi_{\tau} \to 0$,  $\rho_\theta(\omega) \to \tanh(\beta\omega/2)$, i.e.. At finite $\beta/\xi_{\tau}$, $\xi_{\tau}^{-1}$ replaces $\beta^{-1}$ as the infra-red cut-off. A different asymptotic form prevails at  large $\beta \omega$. The fit to the numerical results is shown in the inset of the figure with analytic forms given in the figure caption and reproduced in Eqs. (\ref{gthetaomegak1}), (\ref{gthetaomegak2}).


\begin{thebibliography}{99}

\bibitem{chakra86}
S. Chakravarty, G. L. Ingold, S. Kivelson, and A. Luther, Phys. Rev. Lett. 56, 2303, (1986).
\bibitem{Fisher86}
M. P. A. Fisher, Phys. Rev. Lett. 57, 885 (1986).
\bibitem{expts}
B. G. Orr, H. M. Jaeger, A. M. Goldman, and C. G. Kuper, Phys. Rev. Lett. 56, 378 (1986); 
A. F. Hebard and M. A. Paalanen, Phys. Rev. B 30, 4063 (1984);
N. Mason and A. Kapitulnik Phys. Rev. Lett. 82, 5341 (1999).

\bibitem{cmv}
C. M. Varma, Phys. Rev. B 55, 14554 (1997); {\it ibid.} 73, 155113 (2006); 
M. E. Simon and C. M. Varma, Phys. Rev. Lett., 89, 247003 (2002).

\bibitem{bourges}
P. Bourges and Y. Sidis, Comptes Rendus Physique, 12,461 (2011).

\bibitem{Aji2007}
V. Aji and C. M. Varma, Phys. Rev. Lett. 99, 067003 (2007); Phys. Rev. B 79, 184501 (2009); Phys. Rev. B 82, 174501 (2010).  
 
\bibitem{cmv-afm}
C.M. Varma, arXiv:1502.00577.

\bibitem{Varma1989}
C. M. Varma, P. B. Littlewood, S. Schmitt-Rink, E. Abrahams, and A. E. Ruckenstein, Phys. Rev. Lett. 63, 1996 (1989).

\bibitem{Hohenberg}
P. C. Hohenberg and B. I. Halperin, Rev. Mod. Phys. 49, 435 (1977). 

\bibitem{Beal-Monod}
M. T. B\'{e}al-Monod and Kazumi Maki, Phys. Rev. Lett. 34, 1461 (1975).

\bibitem{Hertz}
J. A. Hertz, Phys. Rev. B 14, 1165 (1976).

\bibitem{Moriya}
T. Moriya, {\it Spin Fluctuations in Itinerant Electron Magnetism} (Springer, Berlin), (1985).

\bibitem{Lohneysen-rev}
H. v. L{\"o}hneysen, A. Rosch, M. Vojta, and P. W\"olfle, Rev. Mod.
Phys. 79, 1015 (2007).

\bibitem{Zhou}
R. Zhou, Z.Li, J. Yang, D.L. Sun, C.T. Lin, and G.-q. Zheng, Nat. Commun.
4, 2265 (2013).

\bibitem{Analytis}
I. M. Hayes, N.P. Breznay, T. Helm, P. Moll, M. Wartenbe, R. D. McDonald, A. Shekhter, and J. G. Analytis,  arXiv:1412.6484. 

\bibitem{Schroeder}
A. Schr\"{o}der, G. Aeppli, E. Bucher, R. Ramazashvili, and P. Coleman, Phys. Rev. Lett. 80, 5623 (1998).

\bibitem{Schroeder2}
 A. Schr\"{o}der, G. Aeppli, R. Coldea, M. Adams, O. Stockert, H.v. L\"{o}hneysen, E. Bucher, R. Ramazashvili, and P. Coleman, Nature (London) 407, 351 (2000). 

\bibitem{Si}
Q. Si, S. Rabello, K. Ingersent, and J. Lleweilun Smith, Nature (London) 413, 804 (2001).

\bibitem{footnote}
In Refs. [\onlinecite{Schroeder, Schroeder2}], the measured imaginary part of the susceptibility $\chi''({\bf q}, \omega)$ at ${\bf q} = {\bf Q}_0$ has been fit to a local criticality form which has a different form than that derived here. However, we find that there is a very good fit to the measured $\chi''({\bf Q}_0, \omega)$ at different temperatures to $\omega/2T$, for $\omega/2T \lesssim 1$, as derived here. The higher frequency parts can only be fit by introducing a cut-off $\omega_c \approx 3 K$.


\bibitem{Berezinskii}
V. L. Berezinskii, Zh. Eksp. Teor. Fiz. 59, 907 (1970).

\bibitem{Kosterlitz}
J. M. Kosterlitz and D. J. Thouless, J. Phys. C 6, 1181 (1973).

\bibitem{Fisher}
M. P. A. Fisher, G. Grinstein, and S. M. Girvin, Phys. Rev. Lett. {\bf 64}, 587 (1990). 

\bibitem{Janke90}
This is similar to the classical 3D anisotropic XY model, which shows a 3D to 2D crossover behavior. See, e.g., W. Janke and T. Matsui, Phys. Rev. B {\bf 42}, 10673 (1990).  

\bibitem{Sudbo}
E. B. Stiansen, I. B. Sperstad, and A. Sudb{\o}, Phys. Rev. B {\bf 85}, 224531 (2012).

\bibitem{Caldeira1983}
A. O. Caldeira and A. J. Leggett, Ann. Phys. (NY) 149, 374 (1983).

\bibitem{Jose}
J. V. Jos\'{e}, L. P. Kadanoff, S. Kirkpatrick, and D. R. Nelson, Phys. Rev. B 16, 1217 (1977).

\bibitem{Villain}
J. Villain, J. Phys. (Paris) 36, 581 (1975)

\bibitem{polyakov}
A. M. Polyakov, Nucl. Phys. B 120, 429 (1977).

\bibitem{Nelson-Kosterlitz} 
D. R. Nelson and J. M. Kosterlitz, Phys. Rev. Lett. 39, 1201 (1977).

\bibitem{Weber1998}
H. Weber and P. Minnhagen, Phys. Rev. B 37, 5986 (1988). 

\bibitem{Bramwell1994}
S. T. Bramwell and P. C. W. Holdsworth, Phys. Rev. B 49, 8811 (1994). 

\bibitem{prevtheor}
S. Chakravarty, G. L. Ingold, S. Kivelson, and G. Zimanyi, Phys. Rev. B 37, 3283 (1988); 
S. Tewari, J. Toner, and S. Chakravarty, Phys. Rev. B 72, 060505(R)  (2005); 
N. Nagaosa, Quantum Field Theory in Condensed Matter Physics (Springer, New York, 1999), Sec. 5.2.

\bibitem{footnote3}
From our numerical calculations, a power law with exponent of 1/8 or smaller cannot be ruled out. This is shown in Appendix \ref{xixxit-powerlaw}.

\bibitem{rafael}
G. Refael, E. Demler, Y. Oreg, and D. S. Fisher, Phys. Rev. B 75, 014522 (2007).

\end{thebibliography}
\end{document}